\documentclass[10pt, review]{elsarticle}
\usepackage[paper=a4paper, top=0.85in, margin=3cm, footskip=2cm, showframe=false]{geometry}
\journal{Arxiv}
\pdfoutput=1
\usepackage{amsmath,amssymb}

\usepackage[utf8x]{inputenc}

\usepackage{textcomp,marvosym}

\usepackage{nameref}
\usepackage[hidelinks]{hyperref}
\hypersetup{
	colorlinks=false,
	pdfborder={0 0 0},
	linkcolor=black,
	citecolor=black,
	filecolor=black,
	urlcolor=black,
}

\usepackage{etoolbox}
\apptocmd{\thebibliography}{\raggedright}{}{}

\usepackage{microtype}
\DisableLigatures[f]{encoding = *, family = * }

\usepackage[table]{xcolor}

\usepackage{array}

\newcolumntype{+}{!{\vrule width 2pt}}

\newlength\savedwidth

\usepackage[aboveskip=1pt,labelfont=bf,labelsep=period,justification=raggedright,singlelinecheck=off]{caption}

\makeatletter
\renewcommand{\@biblabel}[1]{\quad#1.}
\makeatother

\usepackage[nopostdot,nogroupskip,nonumberlist]{glossaries}
\newacronym{pdg}{PDG}{pri\-son\-er's di\-lem\-ma ga\-me}
\newacronym{mturk}{MTurk}{Amazon's Mechanical Turk}
\newacronym{lse}{LSE}{London School of Economics}
\newacronym{brl}{BRL}{Behavioral Research Lab}
\newacronym{egt}{EGT}{evolutionary game theory}
\let\oldgls\gls
\renewcommand{\gls}[1]{\oldgls*{#1}}

\usepackage{cleveref}

\usepackage{graphicx}
\def\figwidth{0.9\textwidth}

\usepackage{booktabs}

\DeclareMathOperator{\e}{e}
\newcommand{\pp}[1]{\left(#1\right)}

\usepackage{lastpage,fancyhdr,graphicx}
\usepackage{epstopdf}
\title{\textbf{Self-interested behaviour as a social norm}}

\begin{document}
\maketitle
\begin{flushleft}
Kamilla Haworth Buchter\textsuperscript{1\Yinyang},
Bjarke Mønsted\textsuperscript{2\Yinyang},
Sune Lehmann\textsuperscript{2*}
\\
\bigskip
\textbf{1} London School of Economics, department of philosophy, logic, and scientific method, London, United Kingdom
\\
\textbf{2} Technical University of Denmark, department of applied mathematics and computer science, 2800 Kgs. Lyngby, Denmark
\\
\bigskip

%
%
\Yinyang These authors contributed equally to this work.

* sljo@dtu.dk

\end{flushleft}
\section*{Abstract}
Language can exert a strong influence on human behaviour.
In experimental studies, it is for example well-known that the framing of an experiment\cite{tversky1981framing} or priming at the beginning of an experiment\cite{molden2014understanding} can alter participants' behaviour.
However, few studies have been conducted to determine why framing or priming specific words can alter people's behaviour\cite{marwell1981economists,gerlach2017games}.
Here, we show that the behaviour of participants in a game-theoretical experiment is driven mainly by social norms\cite{Bicchieri2005}, and that participants' adherence to different social norms is influenced by the exposure to economic terminology.
To explore how these terminology-driven changes impact behavior at the system level, we use established frameworks for modeling collective cooperative be\-ha\-vi\-our\cite{Szabo2007, Lieberman2005}. 
We find that economic terminology induces a behavioural difference which is larg\-er than that caused by financial incentives in the magnitude usually employed in experiments and simulation.
These findings place an increased responsibility on scientists and science communicators, as scientific terminology is increasingly communicated to the general population\cite{davies2008constructing, sugimoto2013scholars, kirby2017changing}.

\section*{Introduction}
We start from the observation that economists tend to exhibit substantially different behaviour in experiments pertaining to game theory and economics\cite{marwell1981economists}.
In particular, those who study, or have studied, economics tend to act more in accordance with the predictions of microeconomic theory, acting to maximize their own profits\cite{gerlach2017games}.
This raises the question why economists behave differently.
Some have proposed a self-selection mechanism, according to which people more inclined to maximize their own profits are also more likely to choose to study economics\cite{carter1991economists}. Others have proposed a learning mechanism where exposure to economic theory is the cause of the behavioural differences\cite{frank1993does}. In the past ten years, studies have repeatedly confirmed the latter mechanism where changes in behaviour are caused by exposure to microeconomic theory\cite{ifcher2018rapid}. However, few studies have engaged with the question of why a learning effect occurs\cite{marwell1981economists,Ferraro2005,cappelen2015social,gerlach2017games}.

Here, we present evidence that engaging with microeconomic terminology inhibits cooperative behaviour in a competitive game setting, and that engaging with an alternative terminology which emphasizes collective, rather than individual, payouts increases cooperative behaviour.
Applying Bicchieri's definition of social norms\cite{Bicchieri2005}, we next show that these terminology-driven changes are caused by social norms and not by alternative effects such as the terminologies enabling participants to understand the experiment or biasing participants to prefer certain behaviours. 
We finally use simulation methods from evolutionary game theory to assess whether observed individual behavioural differences arising from terminology exposures are sufficient to take a collective system across a tipping point to states of complete cooperation or defection in a simulated population.
The findings have impact beyond economics since they suggest that scientific terminology can guide people's behaviour by prompting specific social norms and that behaviour guided by these norms can completely determine the outcome of a collective system.

For the initial experiments, we built an online platform where participants could play 10 rounds of \gls{pdg}.
In a \gls{pdg}, participants are faced with a choice to either \textit{defect} or \textit{cooperate} with another player.
If both players cooperate, their combined payout will be maximized. However, for each individual player, choosing defection will maximize their own payout, regardless of the choice of the other player.
Finally, if both players' defect, they will each receive a smaller payout compared to the payout they get if both players cooperate.
In order to test the behavioural effects of scientific terminology, participants were randomly assigned to one of three categories - \textit{individualist} (I), \textit{collectivist} (C), or \textit{neutral} (N). Before proceeding to the game, participants in all three groups had the structure of the \gls{pdg} explained to them, and were required to correctly answer a series of control questions to ensure they understood how their payout depended on the actions of both players.
In addition, participants in (C) and (I) were introduced to two distinct concepts from microeconomics, and asked to apply them when answering the control questions, and throughout the \gls{pdg}.
Participants in (I) were shown introduced to the microeconomic concept of \textit{rationality} and explained that in game theory it is called \textit{rational} to maximize ones own reward by defecting\cite{mas1995microeconomic}. 
%
In similar fashion, participants in (C) were introduced to the concept of \textit{social optimality}\cite{Binmore1997} and explained that in game theory it is called \textit{optimal} to maximize collective wealth by cooperating.
Players in (N) were not introduced to any concepts and their five control questions emphasized collective and individual gain to the same extent.

Having tested for the behavioural effects of terminologies, we subsequently ran a follow-up experiment to understand the role of social norms in the decision process. 
Here, participants were asked two questions before playing each \gls{pdg}. The first question asked which move they expected the other player to make. The second questions asked which move they thought, the other player expected them to make. 
Our goal was to determine whether there is a connection between participants' expectations and their behaviour and whether the behavioural effects caused by the terminologies are driven by these expectations such that they are caused by adherence to a social norm of cooperation or to a social norm of defection\cite{Bicchieri2005}.

If terminology can substantially change which social norms individuals follow, the question then becomes: What is the effect of terminology at the collective level?
We explore this question through the lens of evolution of cooperative behavior\cite{Axelrod1981, Milinski1987, Trivers1971}.
Specifically we simulate agents embedded in a network and use decision heuristics informed by empirical data on terminology-driven changes in social norms to decide agents' actions in a \gls{pdg} based on their surroundings, following \cite{Nowak1992, Blume1993, Rand2013, Santos2005, Nowak2006}.
We run these simulation experiments on artificial networks commonly used in the literature, as well as networks constructed from real world data, using data from the Copenhagen Networks Study\cite{Stopczynski2014}.

\section*{Terminology influences behaviour}
In order to assess the interplay between terminology and educational background, we recruited participants in the (\gls{brl}) at the \gls{lse} ($n=462$) and on \gls{mturk} ($n=344$).
The participants played a \gls{pdg} with 10 rounds on an 
online platform, through which they were informed they would play against a new participant in each round. In reality, they were playing a computer choosing randomly between defection and cooperation.
The experimental setup was approved by the \gls{lse} Research Ethics Committee.

Among the \gls{brl} participants, $77$ were associated with a degree involving at least two years with economics courses.
The degrees were core degrees from the three departments Economics, Finance, and Accounting, and we refer to participants associated with these degrees as \textit{economists} for short.
Consistent with results from previous \gls{pdg} ex\-peri\-ments\cite{frank1993does,ifcher2018rapid}, we find that economists cooperate less than the remaining \gls{brl} participants (linear regression on number of cooperate moves, $p = .0007$, $n=462$, $t=-3.22$, one-tailed).
Comparing data sources, \gls{brl} participants cooperated significantly less ($p = .048$, $n=810$, $t=-1.66$, one-tailed).
However, this difference disappears when excluding economists from the analysis ($p=.25$, $n=733$, $t=-0.66$, one-tailed).

\begin{figure}[htbp]
	\centering
	\includegraphics[width=\figwidth]{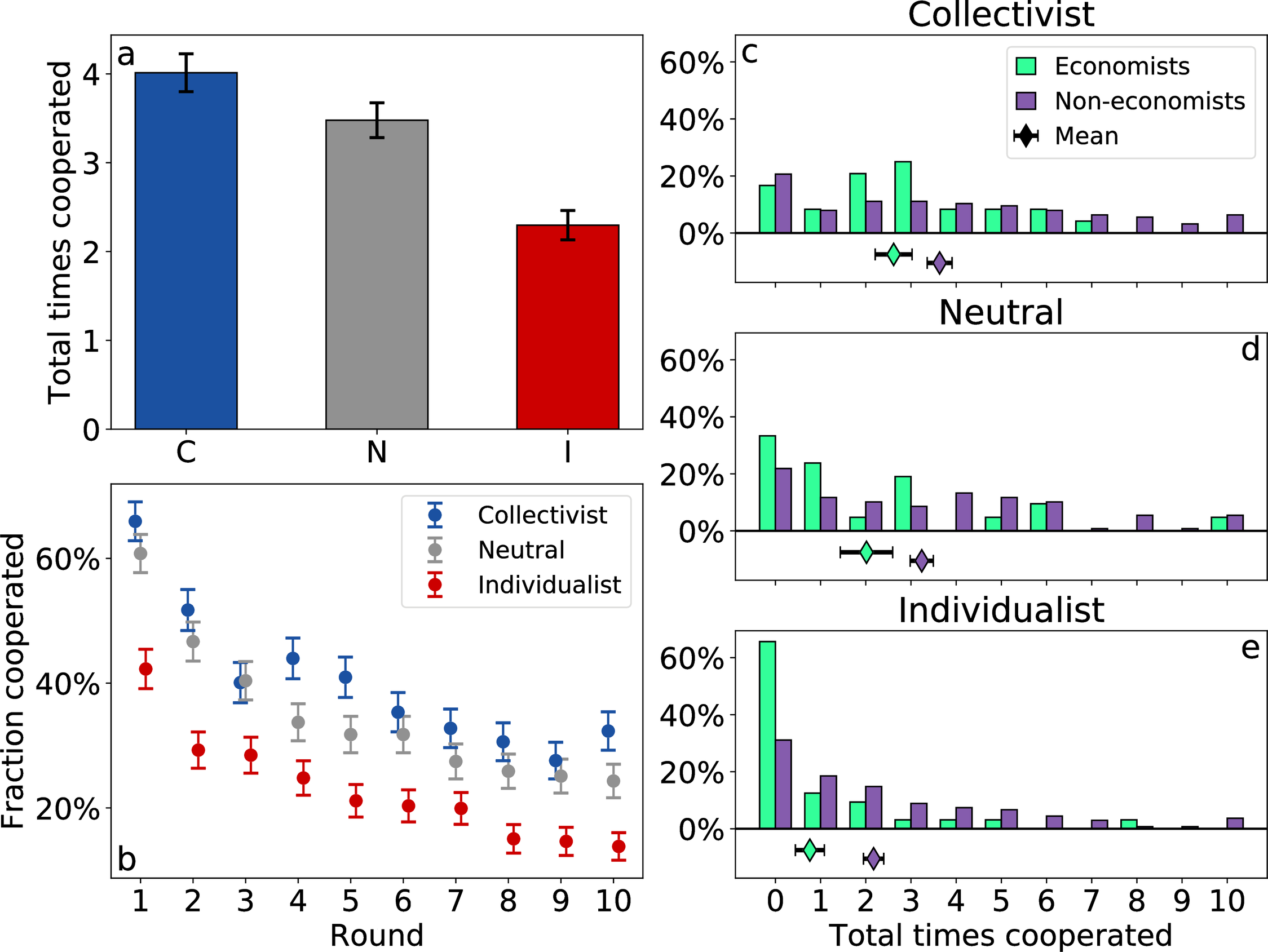}
	\caption{Interaction between terminology and behaviour. \textbf{a} Participants who had the game described in collectivist (C) or individualist (I) terms cooperated and defected more, respectively. \textbf{b} throughout the 10 rounds, participants in all categories became increasingly more likely to defect. Some difference persisted over time, especially individualists defecting more than others. \textbf{c-e}
		Distributions over the number of cooperation moves among participants exposed to the three terminologies, grouped by whether participants had a background in economics. Bars represent the observed frequencies, and the markers represent the mean number of times cooperated. Participants affiliated with an economics-related discipline consistently cooperated fewer times than other participants, but terminology exposure had a similar influence on both groups.
		The mean for each distribution is indicated with black markers. Error bars in all subfigures represent the standard error of the means.}
	\label{fig:behaviour}
\end{figure}

Focusing on the remaining $733$ participants from \gls{brl} and \gls{mturk}, we compare the number of times participants in the collectivist, neutral, and individualist categories defected in the 10 rounds (\cref{fig:behaviour}a).
A Kruskal-Wallis test revealed a significant interaction between terminology and cooperation ($p=6.8\cdot 10^{-9}$, $n=733$, $H=38$). A Conover-Iman post-hoc analysis\cite{Conover1979} showed that, compared with the neutral category, participants exposed to the collectivist terminology cooperated more ($p = .03$, $n=487$, $t=1.89$, , one-sided), while the individualist category defected more ($p=8.2\cdot10^{-6}$, $n=501$, $t=4.34$, one-sided).
Throughout the 10 rounds, participants in all categories became more likely to defect (\cref{fig:behaviour}b).

Considering the \gls{brl} participants, the difference between e\-con\-o\-mists and non-economists is consistent across all categories (\cref{fig:behaviour}c-e).
The magnitude of this behavioural difference between economists and non-economists is comparable to the effect of economic terminology.
In particular, economists exposed to a neutral terminology defected as much as regular participants exposed to the individualist terminology (\cref{fig:behaviour}d-e).

\section*{Behavioural changes are mediated by social norms}
One potential explanation of why terminology influences behaviour, is that 
language impacts the salience of different social norms\cite{Ferraro2005,gerlach2017games}.
We conducted a second experiment on \gls{mturk} ($n=200$) to understand whether social norms provide an explanation of the observed behavioural effects.

We adopt Bicchieri's definition of social norms\cite{Bicchieri2005}, which views a person's behaviour is expressive of a social norm if 1) they are aware that a behavioural rule exists and applies to their situation, and 2) the person's conforming to the rule is contingent on their first and second order beliefs regarding general compliance, i.e. they must generally expect others to comply with the behavioural rule, and believe others to expect them to comply, too.
Note that conflicting norms can exist under this definition\cite{Bicchieri2005}.
The terminologies in the three categories were designed to provide participants with cues that a behavioural rule of cooperation (defection) exists for the collectivist (individualist) terminologies, see methods for details.

To assess beliefs regarding compliance, participants were asked before each round whether they expected the other player to defect or cooperate, and which choice they believed the other player expected of them in turn.
In the following, we will use the phrase 'expecting cooperation (defection)' as shorthand for participants who expect cooperation (defection) of their opponent and believes the same is expected of them.
Note that the two are not exhaustive, as the first and second order beliefs need not align.
If the mechanism through which terminology influences behaviour is social norms of cooperation and defection, we should expect any behavioural effect to be strongly contingent on beliefs about compliance\cite{Bicchieri2009}.

\begin{figure}[htbp]
	\centering
	\includegraphics[width=\figwidth]{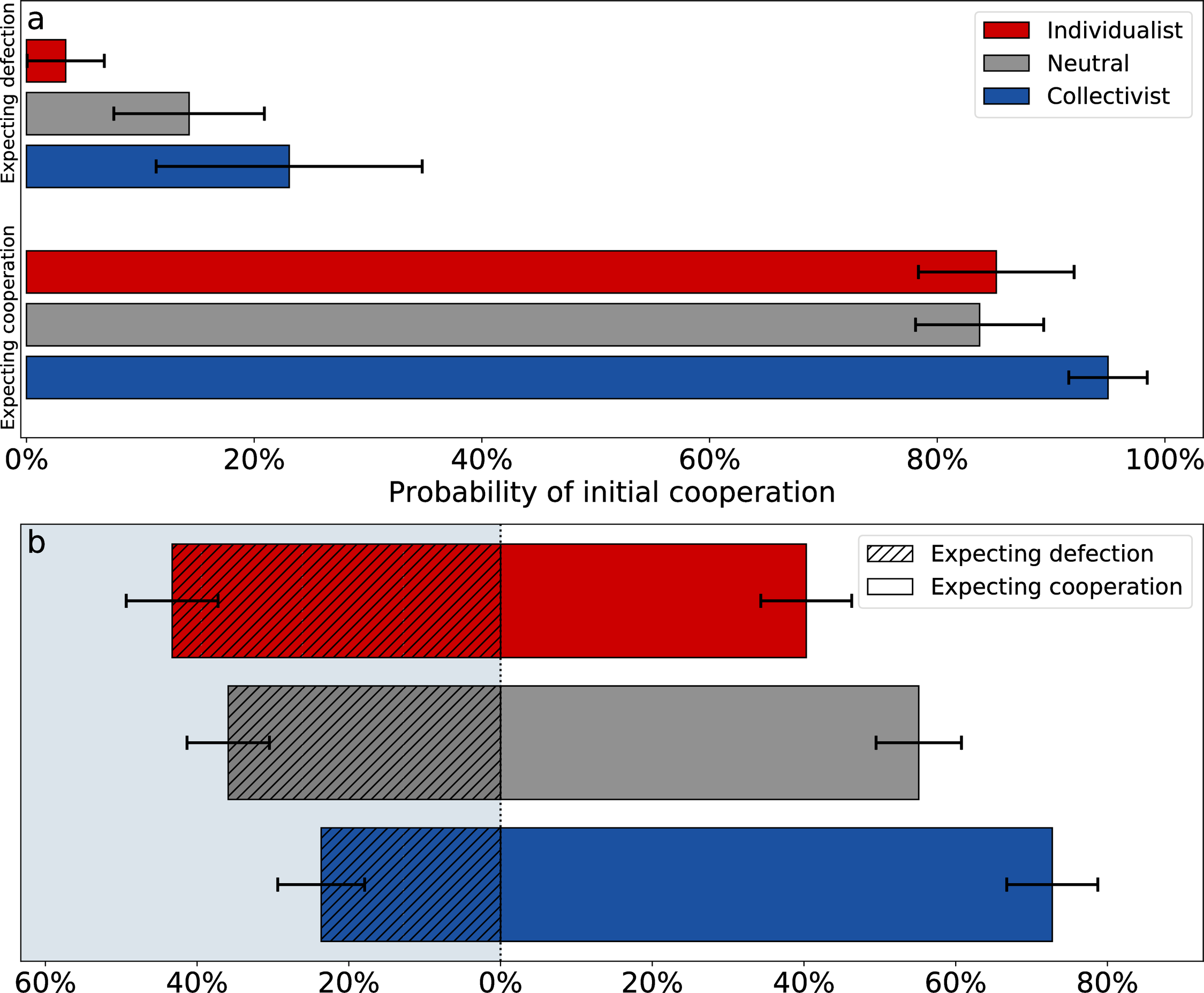}
	\caption{The interplay between expectations and terminology exposure.
		\textbf{a} The percentage of participants choosing to cooperate in each round, grouped by expectations and terminology exposure. When participants expect cooperation, meaning expecting the other player to cooperate and believing the other player expects them to cooperate, they are much more likely to cooperate.
		Similarly, participants who expect defection are much more likely to defect.
		\textbf{b} The percentage of participants exposed to each of the three terminologies (collectivist, neutral, and individualist) who expect cooperation and defection, respectively.
		Participants exposed to the individualist terminology become more likely to expect others to defect, and to believe that others in turn expect defection from them, and less likely to hold the similar beliefs for cooperation. The opposite effect is seen for exposure to the collectivist terms.
		The error bars represent error of the means.}
	\label{fig:norms}
\end{figure}

The results from the second experiment indeed show a strong correlation between expectations and behaviour, as shown in \cref{fig:norms}a.
For example, in the first round, $88.2\%$ of participants expecting cooperation chose to cooperate, whereas $11.2\%$ of those expecting defection did so.
Within the subgroups of participants expecting cooperation and defection, \cref{fig:norms}a shows only a slight difference between participants exposed to the three terminologies.
However, terminology exposure significantly impacts the probability for participants to hold such expectations, as depicted in \cref{fig:norms}b.
Specifically, participants in (C) were significantly more likely to expect cooperation ($p=.02$, $n=133$, $z=2.06$, one-tailed proportional z-test), and participants in (I) significantly less so ($p=.04$, $n = 145$, $z=-1.78$, one-tailed proportional z-test), when compared to the neutral (N) group.
The number of times participants expected cooperation over the ten rounds also varied significantly with terminology exposure ($p=.003$, $n=200$, $H=12$, Kruskal-Wallis test), and similarly for defection expectations ($p=.044$, $n=200$, $H=6.3$).
Excluding participants who responded in a post-experiment survey that they either felt compelled to cooperate or defect more, or suspected that the aim of the experiment was to influence cooperative behaviour, did not significantly influence the results in this or the first experiment (details in methods).

The results visualized in \cref{fig:norms} support the hypothesis that scientific language, specifically microeconomic terminology, can influence people's behaviour by encouraging them to follow different social norms.
We draw this conclusion because the results cannot be adequately explained by the alternative hypotheses. If the terminologies affected behaviour by biasing participants to prefer one action over the other, participants' choices would not depend on their first and second order expectations regarding other par\-ti\-ci\-pants\cite{molden2014understanding}.
Another proposed mechanism for such behavioural effects is that language may shift the underlying utility functions for participants, such that acting e.g. in an altruistic fashion results in a higher utility in spite of the lower payout\cite{cappelen2015social}.
This explanation is also not compatible with the results, as participants were required to answer control questions to ensure their understanding that defection (cooperation) would maximize individual (collective) payout \emph{regardless} of the choice of the other player.
Therefore, if participants were simply maximizing an underlying utility with larger values for individualist or altruistic behaviour, beliefs regarding general compliance should not affect the choices of participants.
For the same reason, it cannot be the case that terminology exposure simply helps participants better understand how to maximize a preexisting individualist or altruist utility function.

\section*{The behavioural effects of terminology can drive simulated collective systems across tipping points}
The experimental results show that terminology can influence individual behaviour in a \gls{pdg} and that this influence can be explained in a social norm framework by a change in beliefs about other people's intentions and expectations.
Having established the impact on social norms at the individual level, we now explore how such terminology-driven shift in norms may impact behavior on the systemic level.
It has been shown that individual behavioural differences caused by social norms may drive a collective system across a tipping point\cite{Nyborg2016} and cause dramatic collective effects.
To investigate whether the observed individual behavioural effect from exposure to microeconomic terminology are sufficient to drive a collective system across a tipping point in our case, we use established methods from agent-based evolutionary game theory\cite{Nowak1992, adami2016evolutionary} to simulate participants exposed to the various terminologies interacting with each other in a network.

We run simulations on a real-world interaction network, obtained from smartphone data from over 700 students in the Copenhagen Networks Study\cite{Stopczynski2014}.
We construct interaction networks in which nodes represents students, and links represent interactions along several channels including text messages, physical proximity (measured by Bluetooth), and Facebook friendships.\cite{Sapiezynski2019}.
We report results from the Facebook friendship network here, and refer to the SI for similar results using the remaining networks, as well as commonly used artificial networks.

Agents in the simulation employ stochastic update heuristics\cite{Wu2007}, spe\-ci\-fi\-cal\-ly a biased logit model\cite{Blume1993},
\begin{equation}
p_c = \frac{1}{1 + \e^{-\beta \mathbf{w}\cdot\mathbf{x}}}, \label{eqn:logit}
\end{equation}
which we fitted to the experimental data.
In \cref{eqn:logit}, $\mathbf{w}$ is a weight vector including bias terms, which were adjusted to the maximum likelihood fit to the experimental data.
$\mathbf{x}$ is a state vector which denotes quantities such as 
the past moves of the agent and their last opponent.

As participants in the three categories (C, N, and I) exhibited quite different behaviours, we repeated this procedure independently for each group, and thus obtained three distinct agent-level decision heuristics.

\begin{figure*}[htbp]
	\centering
	\includegraphics[width=\figwidth]{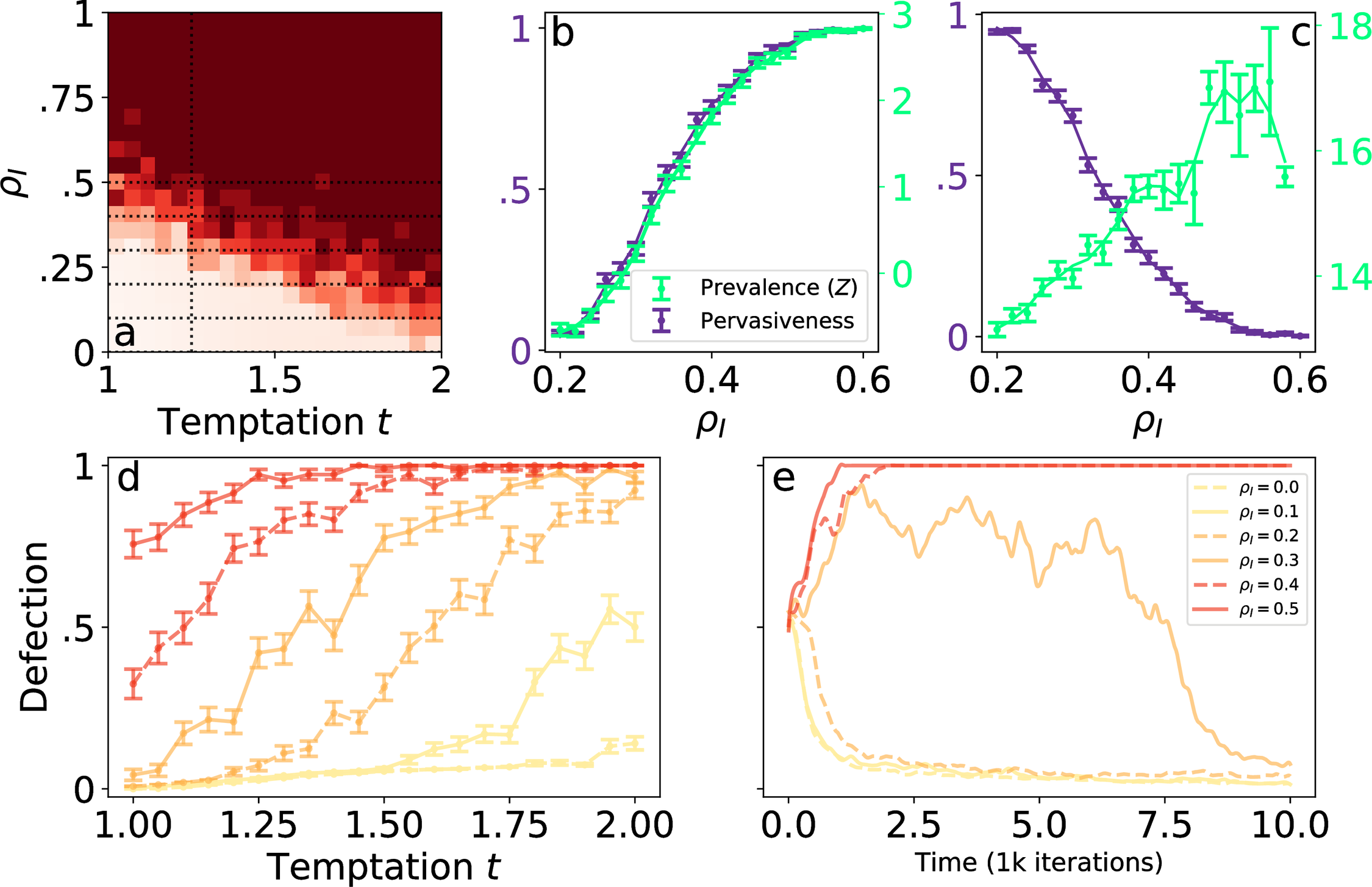}
	\caption{Simulation results for a range of values of the 'temptation to defect' $t$, and the proportion $\rho_I$ of simulated individuals which act similarly to people exposed to individualist, rather than collectivist, terminology.
		\textbf{a} The tendency of cooperation to disappear entirely in simulation, for a range of parameter values. Predictably, this tendency increases with the temptation to defect. However, varying the ratio of collectivists vs. individualists in the simulation has much more pronounced effects, and takes the system across a tipping point, from complete cooperation to complete defection.
		\textbf{b, c} Commonality measures for defection and cooperation, respectively, as a function of the temptation parameter. Global commonality (pervasiveness) increases and decreases, respectively, with temptation. Local commonality (prevalence), however, increases with $\rho_I$ for both strategies, as the network becomes more polarized and cooperators are forced to be more tightly clustered together in order to survive in spite of the growing number of defectors.
		\textbf{d} Defection rates for selected values of $\rho_I$ (indicated with dashed lines in \textbf{a}), averaged over the final 5k iterations of simulations, as a function of temptation.
		\textbf{e} Progression of individual simulations for specific values of $\rho_I$ and $t$ (indicated by the intersections of lines in \textbf{a}). The system generally goes to complete cooperation or defection, but is volatile for values around $\rho_I=.3$. The lines are smoothened using a Savitzky–Golay filter
		with window length 400 and polynomial order 5.
		All error bars represent the error of the means.}
	\label{fig:simulations}
\end{figure*}

In order to assess any emergent phenomena the induced individual behavioural differences might give rise to, we ran a series of simulated repeated \gls{pdg}s.
In order to probe specifically the induced behavioural differences, we subtracted the bias terms from the neutral model from the remaining two, so the resulting models represented the difference in behaviour relative to the neutral category.
Details of this, along with parameter values, fit quality measures, details on alternate models, and visualizations of the resulting heuristics, are available in the SI.
We then defined a parameter $\rho_I$ denoting the fraction of agents in a simulation which used the decision heuristic based on experimental data in the individualist category.
The remaining $1 - \rho_I$ fraction of agents would then act according to the model fitted to data from the collectivist category.
Results from simulations on the real-world Facebook friendship network are illustrated in \cref{fig:simulations}.

In these simulations, results were more sensitive to changes in $\rho_I$ compared to $t$. Scanning the temptation parameter $t$ across the $[1, 2]$ range gradually increases overall defection rates, while increasing $\rho_I$ takes the entire network from complete, or almost complete, cooperation, across a tipping point into a regime of complete defection, regardless of $t$. The proportion of simulations in which cooperators died out entirely is shown for a range of parameter values in \cref{fig:simulations}a.

For selected values of $\rho_I$, the outcomes of simulations for varying temptation values are shown in \cref{fig:simulations}b. For the same values of $\rho_I$, and a single value of $t$, the progression of a single simulation over $10,000$ rounds is displayed in \cref{fig:simulations}c, showing that for high and low $\rho_I$, the system quickly settles to states of complete defection and cooperation, respectively.
For values in between, the system is volatile and can go either way.

While increasing $\rho_I$ leads to decreased cooperation overall, the cooperators that remain tend to be more tightly clustered together.
We can visualize this by dividing the notion of commonality into \textit{pervasiveness} (global) and \textit{prevalence} (local).
We define the pervasiveness of a behaviour (cooperation or defection) simply as the fraction of nodes in a network that partake in that behaviour at the end of a simulation.
Prevalence is defined as the $z$-score of the observed number of e.g. cooperate-cooperate neighbours, compared to a random permutation null model.
\Cref{fig:simulations}d shows the commonality measures for defection as $\rho_I$ is increased, and \cref{fig:simulations}e shows the same for cooperation, showing that cooperators become much more tightly nit together (increasing prevalence) as their numbers dwindle (decreasing pervasiveness).

\section*{Discussion}
Our findings indicate that cooperative behaviour may be significantly influenced by exposure to scientific terminology.
Compared with a neutral group, we found that using different, but equivalent, scientific terms to describe a competitive experiment, we could both amplify and dampen cooperative behaviour.
The terminology which reduced cooperative behaviour is standard microeconomic terminology, and the reduction in cooperative behaviour was comparable to the difference observed between participants enrolled in educations with and without a heavy background in economics.
We saw strong evidence that people's choice to cooperate or defect were in part governed by social norms, as participants were much more likely to elect a move which they expected their opponent to play, and which they believed their opponent in turn expected of them.
The terminologies were able to alter behaviour in our participants, by manipulating these expectations.
We then used simulations to show that the behavioural differences introduced by different terminologies were sufficient to drive a system across a tipping point between states of complete cooperation and defection.

As science is becoming more broadly and popularly disseminated in the population, this places a greater responsibility on science communicators to be understand that scientific language may contain value laden terms which interact with social norms to produce emergent behavioural changes.
In addition, the findings emphasize the need for the experimental behavioural scientist to be aware of how observed behaviours might be influenced by familiarity with relevant scientific terminologies, both from the explanation of the experiment received by participants, and due to popular-scientific communication.
On a grander canvas, our results highlight how, for example, the language of leaders or media -- choosing to focusing on selfish rationality or social cooperation -- may drive real changes in people's behavior.

In future experiments, it would be interesting to look for similar interactions between norms and terminology within other scientific disciplines, as well as to expose groups of participants to the same terminological stimulus to investigate directly the degree to which terms give rise to emergent effects. Finally, further experimentation might attempt to directly influence empirical and normative expectations, by providing participants with direct evidence that a certain behaviour, e.g. cooperation or defection, is more probable from other participants, or that other participants are more likely to expect said behaviour.

\section*{Materials and methods}
\subsection*{Experimental setup and data collection}
The initial experiment on \gls{mturk} was carried out in November 2018.
We used a predefined setting on the \gls{mturk} platform to allow only 'master' workers, with a consistent history of delivering high quality work, to assess the experiment.
Workers were paid a base pay of 2 USD for participating, plus $10\%$ of the points gained throughout the experiment as USD, resulting in a median earning of 3.6 USD.
A total of 344 workers participated.

The experiment was repeated in December 2018, in the \gls{brl} at the \gls{lse}, with 466 participants completing the experiment in batches of up to 20 people over the course of a week.
4 participants were excluded for various reasons - 1 did not have a valid birth year in the laboratory's database, 1 did not identify as either male or female, and 2 had opted not to fill out a short post-experiment questionnaire in which participants could indicate whether they had played a similar game previously.

Here, participants where required to be physically present in the laboratory, playing the \gls{pdg} on computers located in individual booths.
Participants received a base pay of 5 GBP plus another $10\%$ of the points obtained in the experiment in GBP, resulting in a median earning of 6.7 GBP.
Of the 466 participants, 77 associated with a degree in either economics, finance, or accounting, each of which requires at least 2 years with economics courses.

Finally, 200 participants were again recruited on \gls{mturk} in March 2019 to conduct the follow-up experiment in which inquired about participants' expectations regarding the choices of other players, as well as their beliefs about the expectations of others.
The workers were compensated in a similar fashion as in the earlier \gls{mturk} experiment, with a median pay of 2.8 USD.

\begin{figure}[htbp]
	\centering
	\includegraphics[width=0.7\textwidth]{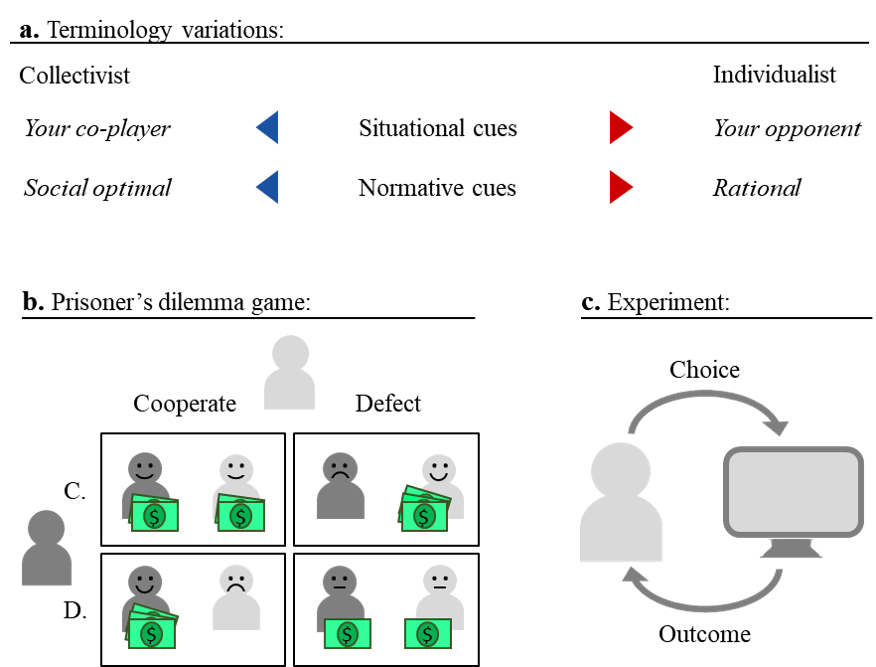}
	\caption{Overview of the experimental setup. \textbf{a} Participants in the collectivist (C) and individualist (I) categories receive situational cues emphasizing competitive or communal aspects of the game. Participants in the neutral (N) category received no normative cues and other participants were referred to as ``the other participant'' to avoid situational cues. \textbf{b} The structure of a prisoner's dilemma game. If players cooperate, their combined payout is maximized. Each individual player will receive a larger playout from defecting rather than cooperating, independently of the other player's action. \textbf{c} Participants answer a series of questions to ensure they understand the game and the terminology being introduced to them. In each round, they choose to cooperate or defect, and the computer presents them with the results of the present round, and asks them a follow-up question involving the introduced terminology to ensure their continued understanding and engagement.}
	\label{fig:sketch}
\end{figure}

Players in the three categories were provided with different \textit{situational cues} in the experimental description by referring to other participants as either \textit{your opponent} (I), \textit{the other participant} (N), or \textit{your co-player} (C). This was intended to highlight competitive or communal aspects of the game (\cref{fig:sketch}a).

Players were not allowed to proceed to the \gls{pdg} before they had answered all five control questions correctly. In this way we ensured that players understood the game and the terminologies we had introduced.
In addition, players would be asked to apply the learned terminology after each of the rounds, to ensure their continued engagement (\cref{fig:sketch}c). More details and screenshots of the platform are provided in the SI.

The additional text and focus in the control questions in (I) and (C) provided players with normative cues, by applying value-laden labels to the defect and cooperation strategies (\cref{fig:sketch}a).

\subsection*{Assessing experimenter bias}
In order to mitigate effects of experimenter bias, participants were asked - after completing the experiment - whether they suspected that a particular hypothesis was being tested.
Participants who responded positively were allowed to describe in a text field what they believed was being tested in the experiment.
We identified responses which indicated that participants either believed that some effect of language on behaviour was being tested, or that a particular action was in some way the right one. 
A summary of the responses is shown in \cref{tab:experimenter_bias}.

\begin{table}[htbp]
	\centering
	\footnotesize
	\begin{tabular}{lccc|ccc|ccc}
		\multicolumn{1}{c}{} & \multicolumn{3}{c}{BRL} & \multicolumn{3}{c}{MTurk 1} & \multicolumn{3}{c}{MTurk 2} \\
		\toprule
		\multicolumn{1}{c}{} & C & N & I & C & N & I & C & N & I \\
		\midrule
		``Language'' & 6 & 0 & 7 & 1 & 1 & 2 & 2 & 0 & 1 \\
		``Defect'' & 13 & 12 & 24 & 2 & 2 & 3 & 0 & 0 & 0 \\
		``Cooperate'' & 8 & 4 & 3 & 0 & 1 & 2 & 1 & 0 & 0 \\
		\bottomrule
	\end{tabular}
	\caption{Overview of the number of people who reported in a post-experiment questionnaire to suspect that various effects were being tested. ``Language'' means the participant stated that they believed the aim of the experiment was to test some form of effect of language on participant behaviour. ``Defect'' and ``cooperate'' refers to participants believing that defection or cooperation was the right strategy to choose.}
	\label{tab:experimenter_bias}
\end{table}

Redoing the statistical analyses while excluding participants from \cref{tab:experimenter_bias}, the main findings remained significant.
The number of times participants cooperated was still different in the three terminology groups ($p=1.4*10{-8}$, $n=660$, $H=36$, Kruskal-Wallis test).
The fraction of participants expecting cooperation (believing their opponent would cooperate, and believing the same was expected of them) in the first round was also higher for the collectivist terminology group ($p = .017$, $n=192$, $z = 2.11$, logistic regression, one-tailed z-test), but not so for defection ($p=.15$, $n=192$, $z=1.03$).
However the total number of times participants expected cooperation/defection across the ten rounds remained significantly higher in the collectivist/individualist exposure groups.

\subsection*{Models and parameter estimation}
\label{sec:heuristic_fitting}
We model the choices of an individual agent as a stochastic function of variables representing information available to the agent. We considered the family of logit dynamics\cite{Blume1993} models, which has previously been used in the context of evolutionary game theory\cite{Wu2007}. We write a general logit dynamics model on the form
\begin{align}
\begin{split}
p_i &= \frac{\e^{\beta\mathbf{w}_i\cdot \mathbf{x}}}{Z}, \\
Z &= \sum_j \e^{\beta\mathbf{w}_j\mathbf{\cdot x}}.  \label{eqn:logit_dynamics}
\end{split}
\end{align}
Here, $\mathbf{x}$ is an input vector representing the information based on which an agent makes their decision, and $\mathbf{w_i}$ is a weight vector that represents the real-valued relative importance of each component of the information $\mathbf{x}$ for deciding upon choice $i$. Hence, one may view
\begin{equation}
G_i(\mathbf{x}) = \mathbf{w}_i\cdot \mathbf{x} \label{eqn:G_weight_thingy}
\end{equation}
as representing the degree to which the available information $\mathbf{x}$ favors a decision of $i$ to the agent. The $\beta$ parameter determines the degree of stochasticity, so that probability distribution arising from the model depends more strongly on the $G_i$ for large values of $\beta$. When $\beta = 0$, the choice is made uniformly at random, independently of the $G_i$, and in the limit where $\beta \rightarrow \infty$, the option $i'$ corresponding to the greatest value of $G_i$ is chosen deterministically, with ties being broken randomly.

To allow for biases in an agent's decision, we introduce into \ref{eqn:logit_dynamics} a bias term $b_i$, which we add to $G_i$. To retain the vector notation for $G_i$ in \ref{eqn:logit_dynamics}, we put the bias term as the first element of the weight vector, $w_i^{(0)} = b_i$, and let the corresponding element of the input vector be unity $x^{(0)} = 1$.

We denote the possible choice in a given round of the \gls{pdg} $c$, for 'cooperate', and $d$ for 'defect'. The probability distribution over the choices is is then given by:
\begin{align}
\begin{split}
p_c &= \frac{\e^{\beta\mathbf{w}_c\cdot \mathbf{x}}}{\e^{\beta\mathbf{w}_c\cdot \mathbf{x}} + \e^{\beta\mathbf{w}_d\cdot \mathbf{x}}},  \\
p_d &= \frac{\e^{\beta\mathbf{w}_d\cdot \mathbf{x}}}{\e^{\beta\mathbf{w}_c\cdot \mathbf{x}} + \e^{\beta\mathbf{w}_d\cdot \mathbf{x}}}.
\end{split}
h\end{align}
Due to normalization, this may be simplified as
\begin{equation}
\begin{split}
p_d &= \frac{1}{1 + \e^{-\beta \mathbf{w}\cdot\mathbf{x}}}, \\
p_c &= 1 - p_d, \\
\end{split} \label{eqn:deltamodel}
\end{equation}
where
\begin{equation}
\mathbf{w} = \mathbf{w}_d - \mathbf{w}_c
\end{equation}
Based on this, we tried fitting several different models to the data:
\begin{enumerate}
	\item One model in which the state vector $\mathbf{x}$ of eq. \ref{eqn:G_weight_thingy} contains only information on defecting behavior, i.e. indicator variables for whether the player and their opponent defected in the previous round, and the payouts obtained by the player and the random neighbor, in the event they defected. For instance, the indicator variable denoting whether the player defected in the previous round would be $\delta_{p, d}$, i.e. a Kronecker delta taking the value $1$ if the player defected in the previous round and $0$ otherwise, and the payout variable would be $\delta_{p, d}\cdot f_p$, i.e. the player's payout from the previous round if they defected, and zero otherwise. 
	\item A similar model, but allowing separate values and indicator variables for cooperation. \label{model:DeltaTwo}
	\item A model as the above, but allowing the bias term to depend on the previous action taken by the player and their opponent. \label{model:QuadDelta}
\end{enumerate}
The free parameters of each model were then fitted using the CO\-BY\-LA optimization method\cite{Powell1994} to minimize the negative log-likelihood of model. Some constraints were imposed upon the parameters to avoid performance degeneracies in the parameter space - $\beta$ was constrained to positive values, and the norm $|\mathbf{w}|$ of the weight vector was set to unity. The fitting procedure was repeated 10 times with parameter vectors randomly initialized in each run to mitigate the problem of local optima. The majority of runs converged and resulted in very similar negative log-likelihoods, with a few outliers at greater values, confirming the necessity of multiple runs of the fitting procedure. For the majority of runs which both converged and had similar likelihoods, the parameter vectors returned by the algorithm were closely clustered together. This was not the case when the aforementioned constraints were omitted, indicating that the constraints were indeed necessary to remove parameter space degeneracies. Performance metrics for the best fits for each model are summarized in table \ref{tab:fitting_summary}.

In addition to the models described above, two additional types of models were considered. One such type of models was similar to models \ref{model:DeltaTwo} and \ref{model:QuadDelta} above, but instead of incorporating a bias term in \ref{eqn:G_weight_thingy} the bias would be outside the exponential function so $p_d \propto \alpha \e^{\beta \mathbf{w}_d\cdot \mathbf{x}}$, turning \ref{eqn:deltamodel} into $\frac{1}{1 + C \e^{-\beta \mathbf{w}\cdot\mathbf{x}}}$, where $C = \frac{1 - \alpha}{\alpha}$. However, this resulted in the same values of the performance metrics as models \ref{model:DeltaTwo} and \ref{model:QuadDelta}. We also investigated models which took history from the previous two rounds, rather than just one, for the participants into account. This, however, resulted in a slightly worse fit to data, as well as higher model complexities. In addition, we tried a model that also explicitly took into account the $T$ parameter from the payout matrix. This slightly increased model likelihood but decreased the AIC score due to the additional model complexity. For this reason, we proceed with model 3, without the $T$ parameter.

\begin{table}
	\centering
	\footnotesize
	\begin{tabular}{ccccccc}
		\toprule
		Model & $n$ & $-\ln\mathcal{L}(\mathbf{w})$ & Accuracy & AIC & $F1$ \\ 
		\midrule
		1 & 6 & 3011 & .76 & 6059 & .83  \\ 
		2 & 10 & 2777 & .81 & 5614 & .87  \\ 
		3 & 9 & 2725 & .86 & 5503 & .90 \\
		\bottomrule
	\end{tabular}
	\caption{Summary of various performance metrics for the models. The table displays the number of model parameters $n$, and the negative log-likelihood $-\ln\mathcal{L}(\mathbf{w})$ of each model along with its accuracy. As more complex models would be expected to fit any data better, we also provide the Akaike Information Criterion (AIC) score\cite{Akaike1974}. As the data are unbalanced (with many more choice to defect than to cooperate), we also provide the $F1$ score for the models.}
	\label{tab:fitting_summary}
\end{table}

The state vector $\mathbf{x}$ consists of the following components: An indicator of the immediate game history $H$ available to the player, i.e. of whether they and their opponent cooperated or defected in the previous round, as well as the payouts they, and a random person in their neighborhood, received from cooperating and defecting in the previous round. $\mathbf{x}$ may be written as
\begin{equation}
\mathbf{x} = \pp{\delta_{H, cc},
	\delta_{H, cd},
	\delta_{H, dc},
	\delta_{H, dd},
	\delta_{s, d} p_s,
	\delta_{s, c} p_s,
	\delta_{n, d} p_n,
	\delta_{n, c} p_n
}',
\end{equation}
where $\delta_{H, ij}$ is a Kronecker delta which is $1$ if the player and their opponent played strategies $i$ and $j$, respectively, in the previous round. Similarly, $s$ and $n$ are used as indices of the player themselves and the random neighbor, respectively, with $p$ denoting payout.

\subsection*{Parameter adjustments for simulations}
To account for the fact that most simulation approaches use a weak \gls{pdg}, the model requires a few adjustments after the fitting procedure. First, the fitted heuristics display a very strong bias towards defection, possibly because participants in the experiment played a strong prisoner's dilemma game, with a game matrix given by $T = t, R = 2, P = 1, S = 0$, with $t$ lying in $(2, 4)$, whereas in our simulations, in order to align with the literature, we use $T=t, R=1, P=S=0$, with $t$ in $[1, 2]$. Second, the stochasticity in the fitted model leads to low stability in clusters of similarly acting agents, and negates interesting network effects\cite{Gracia-Lazaro2012}. Third, in accordance with other literature finding that experimentally determined neighbor influence is quite low\cite{Grujic2010}, which adversely affects simulations\cite{Gracia-Lazaro2012}.

The latter obstacle we overcome simply by fixing the weights representing the impact from neighbor payouts on an individual's choice to the same as the weights for the individual's own payouts. This matches well with the literature, in which every individual heuristic we encountered also treated payouts for the individual in question and their neighbors on equal footing.
The problem of stability we mitigated by enforcing a rule that if an individual seeking to update their strategy, and the randomly selected individual neighbor with whom they compared strategies and payouts, had both followed the same strategy, the agent would deterministically choose that strategy.
Finally, to compensate for the increased incentive to defect in the strong vs. weak prisoner's dilemma game, we shifted the bias terms so biases for the neutral data were at zero, while retaining differences between bias terms for the three terminologies.
This adjustment was performed in the following way: From $\beta$ and the weight vectors $w$ (which include the bias terms) in \cref{eqn:deltamodel}, a vector of 'absolute weights' $\mathbf{V}$ are computed as $\mathbf{V} = \beta\cdot \mathbf{w}$.
These are equivalent to the weights in e.g. a normal logistic regression model. Notice that, as we've used the constraint $\mathbf{w}$ is $L^2$-normalized, we have $\beta = |\mathbf{V}|$

For a given terminology exposure $i$, the corresponding vector $\mathbf{V}_i$ may be thought of as a concatenation of the weight vector $\mathbf{v}_i$ and a vector of biases $\mathbf{b}_i$, i.e. $\mathbf{V}_i = \mathbf{v}_i\oplus \mathbf{b}_i$.
This vector is then offset by the biases from the neutral terminology, i.e.
\begin{equation}
\mathbf{V}_i' \leftarrow \mathbf{V}_i - \mathbf{0}\oplus\mathbf{b}_n,
\end{equation}
where $\mathbf{b}_n$ represents the biases in the neutral model. We may rewrite this as $\mathbf{V}' = \beta'\mathbf{w}'$ for consistency with previous notation. The parameters for the three models (C, N, and I) after this transformation are displayed in \cref{tab:all_parameters}

\begin{table}
	\centering
	\footnotesize
	\begin{tabular}{cccc|ccc}
		& \multicolumn{3}{c}{Adjusted} & \multicolumn{3}{c}{Raw fits} \\
		\toprule
		
		& C & N & I & C & N & I \\ 
		\midrule
		$w_{sd}$ & .23 & .26 & .22 & .31 & .47 & .34 \\ 
		$w_{sc}$ & .21 & .24 & .17 & .28 & .36 & .31 \\ 
		$w_{nd}$ & .23 & .26 & .22 & .017 & -.01 & .015 \\ 
		$w_{nc}$ & .21 & .24 & .17 & -.042 & -.03 & -.035 \\ 
		$b_{cc}$ & -.11 & 0 & .054 & -.81 & -.78 & -.75 \\ 
		$b_{cd}$ & -.049 & 0 & .1 & -.42 & -.42 & -.25 \\ 
		$b_{dc}$ & -.033 & 0 & .054 & .00 & .052 & .17 \\ 
		$b_{dd}$ & -.046 & 0 & -.0072 & -.02 & .047 & .037 \\ 
		$\beta$  & 4.6 & 3.9 & 5.7 & 3.5  & 3.0  & 2.7  \\
		\bottomrule
	\end{tabular}
	\caption{Parameter values for the agent logit model. The rightmost columns contain the values obtained directly by fitting to experimental data, and the leftmost columns show the adjusted values.}
	\label{tab:all_parameters}
\end{table}

\begin{figure}
	\centering
	\includegraphics[width=0.95\linewidth]{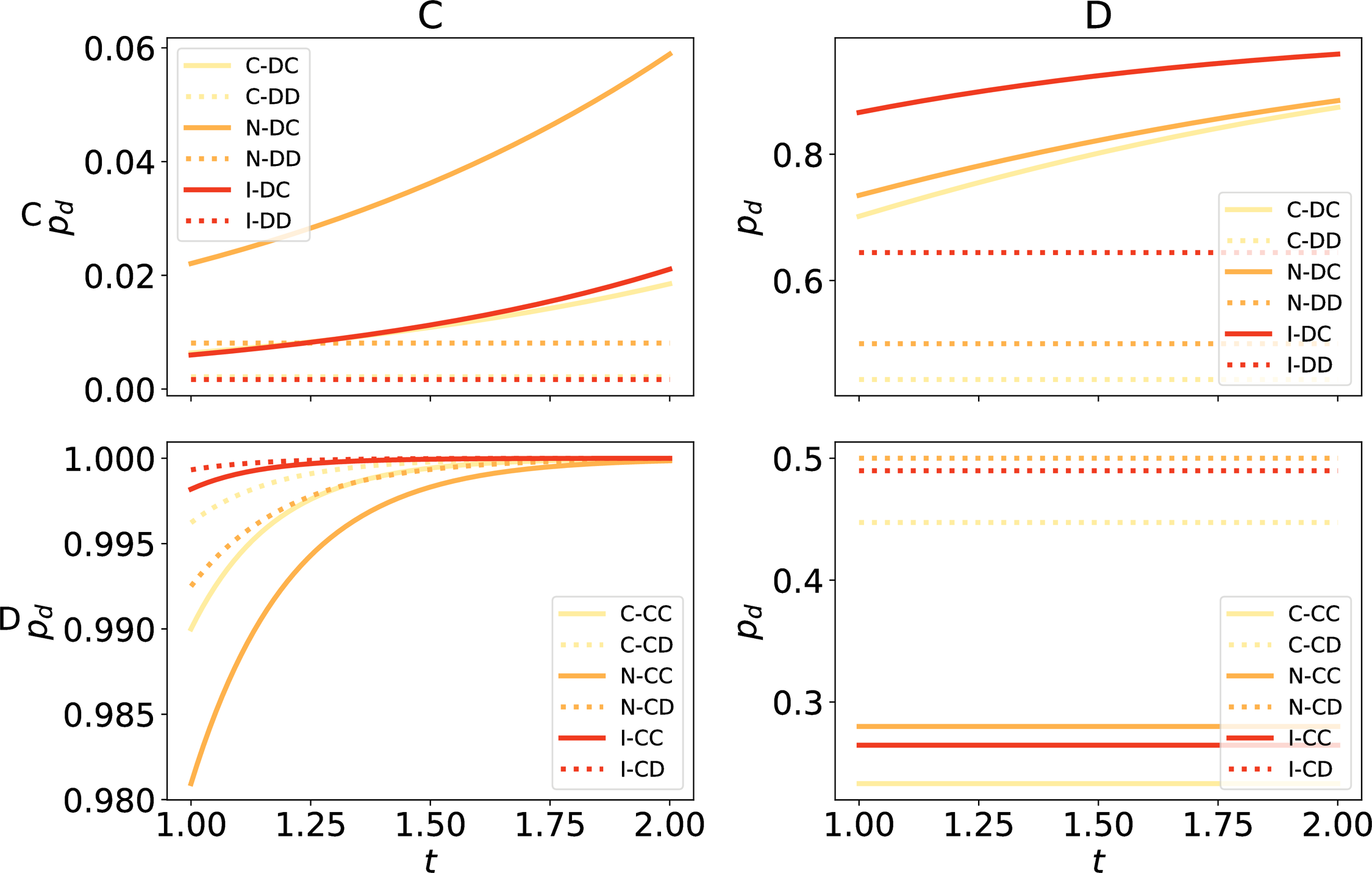}
	\caption{Visual depiction of the agent heuristics with parameters displayed in \cref{tab:all_parameters}. Each plot corresponds to the outcome of a round based on which a given agent is making a decision to potentially update their strategy. Rows correspond to the agent's previous strategy, and columns to their opponent's strategy, so each plot corresponds to the corresponding cell in the game matrix. Each line corresponds to a terminology (C, N, or I), and the strategies chosen by the agent's neighbor, and the neighbor's opponent in the previous round. For example, the label "N-DC" corresponds to a model based on the neutral terminology, and the situation where the agent's neighbor defected while their opponent cooperated. The x-axis represents the $t$ parameter, and the y-axis the probability of the agent defecting $p_d$. Note that we do not show lines for situations where the agent and neighbor played the same strategy, as the agent retains their strategy in that case.}
	\label{fig:heuristicslab}
\end{figure}

The models given by the parameters in \cref{tab:all_parameters} are visualized in \cref{fig:heuristicslab}.

In the main paper, we investigate the effects of terminologies by running simulations in which varying fractions of the individual agents employ the decision heuristics based on the collectivist and individualist terminologies, respectively. We probe this through the parameter $\rho_I$, which denotes the fraction of agents that are randomly assigned to follow the model based on the individualist terminology, labeled 'I' in \cref{fig:heuristicslab}, whereas the remaining $1 - \rho_I$ follow the model based on the collectivist terminology. Hence, values of $\rho_I = 0$ and $\rho_I = 1$ correspond to 'pure' systems in which every agent employs the heuristics from the collectivist and individualist terminologies, respectively, whereas intermediate values correspond to 'mixed' systems in which both groups of agent coexist.

We consider the interactions between terminologies, as expressed by $\rho_I$, and the 'temptation to defect' parameter $t$, and a range of quantities, such as the fraction of agents defecting, the mean payouts for all nodes in the network, a 'pairing measure' capturing to which degree cooperating agents are connected to fellow cooperators at a disproportional rate, etc. As agents embedded in social networks are known to exhibit a high degree of homophily in terms of communication and media consumption\cite{Halberstam2016,Hermida2012,Dvir-Gvirsman2017}, we also investigate the effects of increasing the terminology homophily by assigning terminologies in way which makes agents exposed to similar terminologies more likely to be connected.

We present a brief overview and explanation of these quantities, and give summarize exploratory analyses of their interplay with networks structure and clustering in the SI.

\section*{Acknowledgments}
The authors are grateful to Jason McKenzie Alexander for his insightful comments, to the employees and volunteers at the behavioural research lab at LSE, and to Erik Mohlin, Marco Islam, and Alexandros Rigos for helpful discussions.

\bibliography{references}

\pagebreak
\begin{center}
	\Huge Self-interested behaviour as a social norm\\
	\vspace*{-5mm}\mbox{}\\
	\huge \textit{Supplementary Information Appendix}\\
	\vspace*{2mm}\mbox{}\\
	\large Kamilla Buchter, Bjarke Mønsted, \& Sune Lehmann
\end{center}
\pagebreak

\section*{Experimental setup}
\subsection*{Design of the first experiment}
The experiment had four parts.
First, participants were presented with the title ``A Choice Experiment" and read a general description of the experiment.
Second, they were asked to read a description of the game and answer five control questions which ensured that they understood the rules of the game.
Participants could not proceed to the game before all five control questions were answered correctly.
Third, the participants played ten rounds of PDG. After each round, participants were informed about the choice made by their opponent, their own pay-off from the game, and the choice and pay-off of another random participant.\footnote{The information about the random participant was generated by the computer at random.} The participants were then asked a follow-up question about the random participant in order to make them engage with this information.
The participants could not continue to the next round of PDG before they had answered the follow-up question correctly.
Finally, after playing the ten rounds, participants were asked to state whether they had played this type of game before, and if they had guessed the hypothesis tested in the study.
The experiment took less than 15 minutes to complete.

\begin{figure}[ht]
	\centering
	\includegraphics[width=0.5\textwidth]{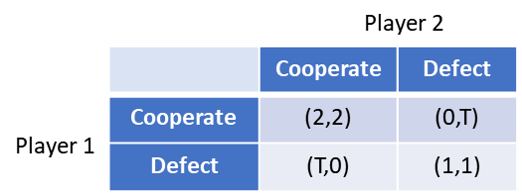}
	\caption{\textbf{Structure of the PDG played in the experiment.} The numbers indicate the points that players can win, where $T\in (2,4)$ is chosen at random for each player. The value of $T$ for each participant remains the same throughout the experiment.}
	\label{fig:experimentPD}
\end{figure}

The PDG played by the participants had the pay-off structure $T>R>P>S$ where $S=0$, $P=1$, $R=2$, and $T$ was selected uniformly at random from the interval $(2,4)$. The two strategies were called \textit{cooperate} and \textit{defect}. The pay-off structure is depicted in table \ref{fig:experimentPD}. 
The entire experimental set-up can be seen at: http://ahura.herokuapp.com/.\footnote{To go to the experiment, enter a random sign in the field for an identification code, press ``I agree - continue to study'', and press ``submit''.}

To test the effect of microeconomics textbooks terminology, participants were randomly allocated to one of three categories in the experiment. The control category used a neutral terminology and did not introduce a microeconomics concept. The second category used an individualist terminology and asked participants to read a text excerpt stating that in game theory the word \textit{rational} is used to denote the strategy of defecting. The third category used a collectivist terminology and asked participants to read a text excerpt stating that in game theory the word \textit{optimal} is used to denote the strategy of cooperation. The two text excerpts were designed to be similar in their formulation, so that the only difference is whether they apply a positively laden word to the strategy of defection or to the strategy of cooperation. Both words relate to concepts used in microeconomics, though rational behaviour is more widely used than socially optimal outcomes.\footnote{Notice that the texts presented in the experiment provide a somewhat simplified version of game theory since rationality and social optimality is concerned with utilities rather than monetary gains. Thus, the concepts only apply to the experiment under the assumption that participants will increase their utility by increasing their wealth.} Next, we provide a detailed account of the stimuli used in each of the three categories.

\subsubsection{Neutral terminology}
The first category in the experiment used a \textit{neutral terminology} in order to provide a control version to which the other two terminologies can be compared. In this category, participants were greeted with the sentence ``Welcome! You are about to take part in a study on how we make strategic choices.'' Further, \textit{the other participant} was used to describe other players in the experiment.
The category did not introduce a microeconomics concept and the control questions pointed both to dominant strategies and to the benefit of cooperation without any normative wording:
\begin{enumerate}
	\item If, in a given round, the other participant and you both play `cooperate', how many points do you receive?
	\item In a given round, you choose `defect' and receive 1 point. Which strategy did the other participant choose?
	\item If the other participant plays `defect' in a given round, which strategy should you choose to ensure that you get the greatest possible number of points?
	\item If the other participant plays `cooperate' in a given round, which strategy should you choose to ensure that the two of you receive the greatest possible total number of points? Hint: for each strategy combination, sum the payoffs you and the other participant will receive.
	\item Assume that in a given round you choose `defect' and receive $T$. Which strategy did the other participant choose?
\end{enumerate}
Finally the follow-up question after each round said:
\begin{itemize}
	\item A random participant played [cooperate/defect] in the previous round, and received a payout of [$T$/2/1/0]. \\
	Which strategy did that player's opponent choose?
\end{itemize}
This ensured that participants engaged with the information they received about the random participant's game.

\subsubsection{Individualist terminology}
The second category of the experiment used an \textit{individualist terminology} to mirror the language of standard microeconomics textbooks. The terminology was introduced through four changes to the experiment.

First, the title of the experiment was accompanied by a small subtitle ``A study on rationality'' and participants were greeted with the sentence ``Welcome! You are about to take part in a study on rationality.'' Further, \textit{your opponent} was used to describe the other players. These situational cues were supposed to indicate to the participants that they were in a competitive situation.

Second, participants were asked to read a short text introducing the microeconomics concept of \textit{rational} to describe the choice of defecting in the game. The text was:
\begin{quote}
	A concept of particular interest in this study is the notion of \textbf{rationality}. In game theory, we say that it is \textbf{rational} for a player to choose a strategy, if the strategy is guaranteed to result in a greater payoff to the player, regardless of which strategy their opponent plays. Conversely, we say that it is \textbf{irrational} for a player to choose a strategy that does not guarantee the highest possible payoff (regardless of what the other player chooses), if a strategy that does so is available. The following contains a few control questions to ensure that you understand these concepts and their relation to the game.
\end{quote}

Third, control questions 3-5 were changed in order to ensure that participants understood the concept of \textit{rationality} and knew how to apply it. The three control questions were:
\begin{enumerate}
	\setcounter{enumi}{2}
	\item If your opponent plays `cooperate' in a given round, which strategy should you choose to ensure that you get the greatest possible number of points?
	\item If your opponent plays `defect' in a given round, which strategy should you choose to ensure that you get the greatest possible number of points?
	\item Given your answers to the above, how would the `defect' strategy be classified according to game theory?
\end{enumerate}

Finally, the follow-up question after each round of the game was changed:
\begin{itemize}
	\item A random participant played [cooperate/defect] in the previous round, and received a payout of [$T$/2/1/0]. \\
	How does game theory categorize this strategy?
\end{itemize}
The participants could either answer \textit{rational} or \textit{irrational}. This change ensured that participants engaged with the information about the random participant and that they used the microeconomics terminology throughout the experiment.

\subsubsection{Collectivist terminology}
The third category used a \textit{collectivist terminology}. The collectivist terminology was designed to mirror the individualist terminology by having a parallel sentence structure. The terminology was introduced through four changes to the experiment.

First, the title of the experiment was accompanied by a small subtitle ``A study on cooperation'' and participants were greeted with the sentence ``Welcome! You are about to take part in a study on cooperation.'' Further, \textit{your co-player} was used to describe the other participants in the experiment. These changes were intended to provide the participants with a situational cue that they were in a cooperative situation.

Second, participants were asked to read a short text introducing the concept of \textit{optimal} to describe the choice of cooperating in the game:
\begin{quote}
	A concept of particular interest in this study is the notion of \textbf{social optimality}. In game theory, we say that an outcome is socially optimal if it results in the largest overall payoff and if no one can be made better of without making someone else worse off. We call a strategy that can lead to a socially optimal outcome \textbf{optimal}. Conversely, we call a strategy \textbf{suboptimal} if it cannot lead to a socially optimal outcome. The following contains a few control questions to ensure that you understand these concepts and their relation to the game.
\end{quote}

Third, control questions 3-5 were changed to ensure that participants understood the concept of optimality and knew how to apply it:
\begin{enumerate}
	\setcounter{enumi}{2}
	\item If your co-player plays `cooperate' in a given round, which strategy should you choose to ensure that the two of you receive the greatest possible total number of points (i.e. which choice maximizes the sum of the points that you and your co-player receive)?
	\item If your co-player plays `defect' in a given round, which strategy should you choose to ensure that the two of you receive the greatest possible total number of points? (i.e. which choice maximizes the sum of the points that you and your co-player receive?)
	\item Given the above how may we classify the role of the `cooperate' strategy in increasing overall wealth?
\end{enumerate}

Finally, the follow-up question was changed to:
\begin{itemize}
	\item A random participant played [cooperate/defect] in the previous round, and received a payout of [$T$/2/1/0]. \\
	Which of the following best describes this strategy choice?
\end{itemize}
The participants could either answer \textit{optimal} or \textit{suboptimal}. 

\subsection*{Design of the second experiment}
In order to ensure that the situation in the second experiment is comparable to the situation in the first experiment, we used the same experimental design as reported above. However, we made one change to the experiment.
In each round of the PDG, participants were asked two additional questions, before they indicated which strategy they wanted to play. The first question was designed to ask about the participants' empirical expectations, while the second question was designed to ask about participants' normative expectations.
The questions were:
\begin{itemize}
	\item Which strategy do you think the other participant will choose?
	\item We also ask the other participant which strategy they think \textbf{you} will choose. What do you think the other participant answers?
\end{itemize}
The participants could answer \textit{cooperate} or \textit{defect} to each question. 
For participants in the collectivist category in the experiment, ``the other participant'' was changed to ``your co-player''. For participants in the individualist category, it was changed to ``your opponent''.

\section*{Interplay between heuristics and network}
\label{sec:heuristics_and_networks}
This section provides an overview of the classes of real and artificial networks we considered for analyses, as well as a range of possible update heuristics for the agents embedded in the simulations. As simulations behave differently in each of the relatively large number of combinations of network types and update heuristics, we provide an overview here along with some qualitative reasons for our choice of focus in the main paper.

The networks considered fall in one of two categories - real, and artificial, i.e. constructed using real-world data, and constructed computationally, starting from a small set of simple rules. The artificial networks under consideration are:
\begin{itemize}
	\item 	A simple 2-dimensional square lattice (SL), such as the one considered in Nowak and May's famous 1992 paper on evolutionary games\cite{Nowak1992}. In this network, each node is connected only to its 4 neighbours - north, south, east, and west, with periodic boundary conditions.
	\item An Erdos–Rényi (ER), in which every pair of nodes $(u, v)$ are randomly connected, each with an independent probability chosen as $1\%$.
	\item A Barási-Albert (BA) scale-free network\cite{Barabasi1999} constructed by starting with a set number $m$ of interconnected nodes, and then grown using a preferential attachment scheme in which each new node is attached to $m$ existing nodes with probabilities proportional to the degrees of the nodes. In the literature, we encountered results from simulations on BA networks with parameter choices ranging from $m=4$\cite{Santos2005} to to $m=8$\cite{Wu2007}, leading us to use $m=6$.
\end{itemize}
All networks mentioned in the above were constructed with a size of $n=625$ nodes.

\begin{figure}
	\centering
	\includegraphics[width=\figwidth]{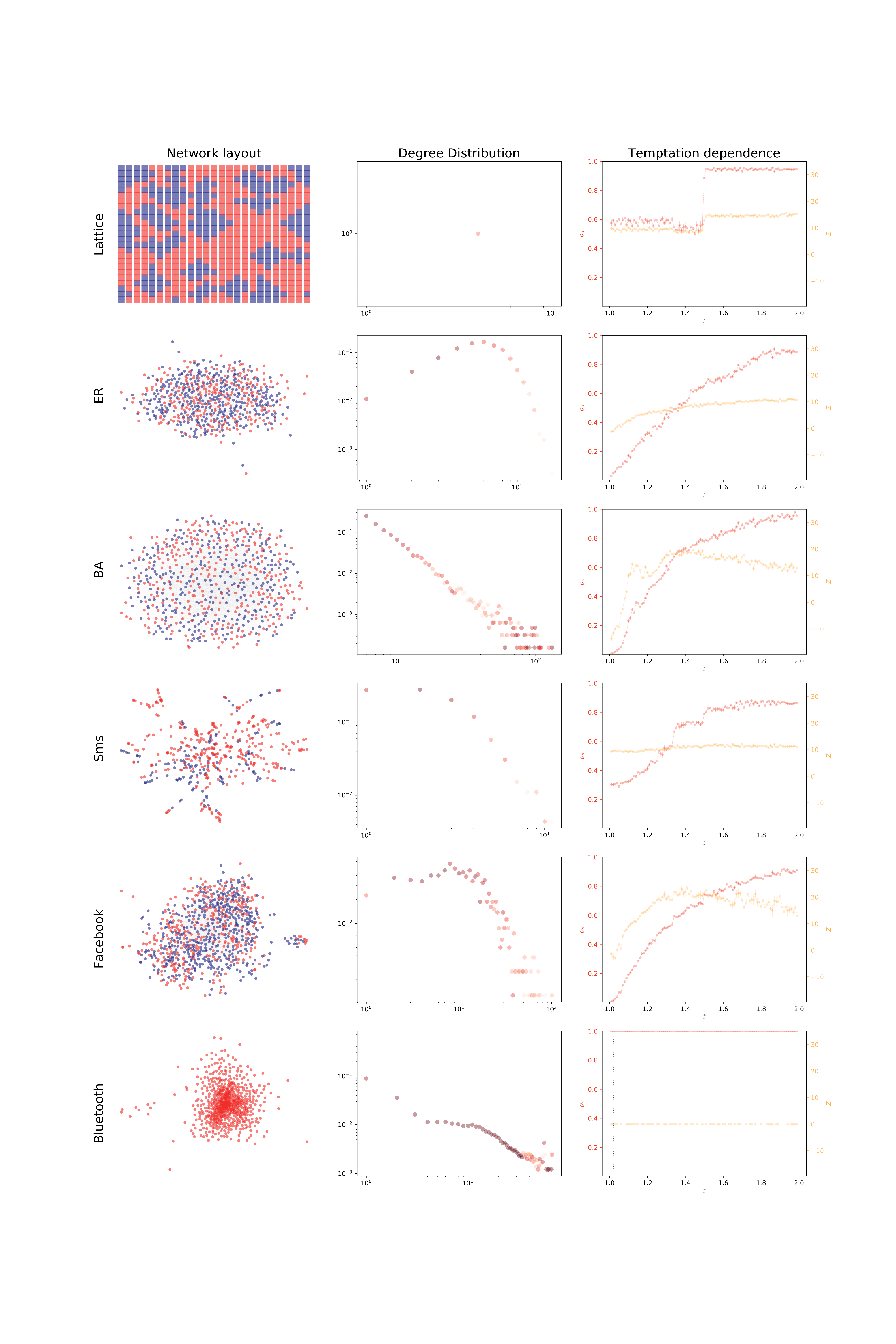}
	\caption{summary of the local maximum heuristic (\ref{it:heuristic_localmax}).}
	\label{fig:nc_localmax}
\end{figure}

\begin{figure}
	\centering
	\includegraphics[width=\figwidth]{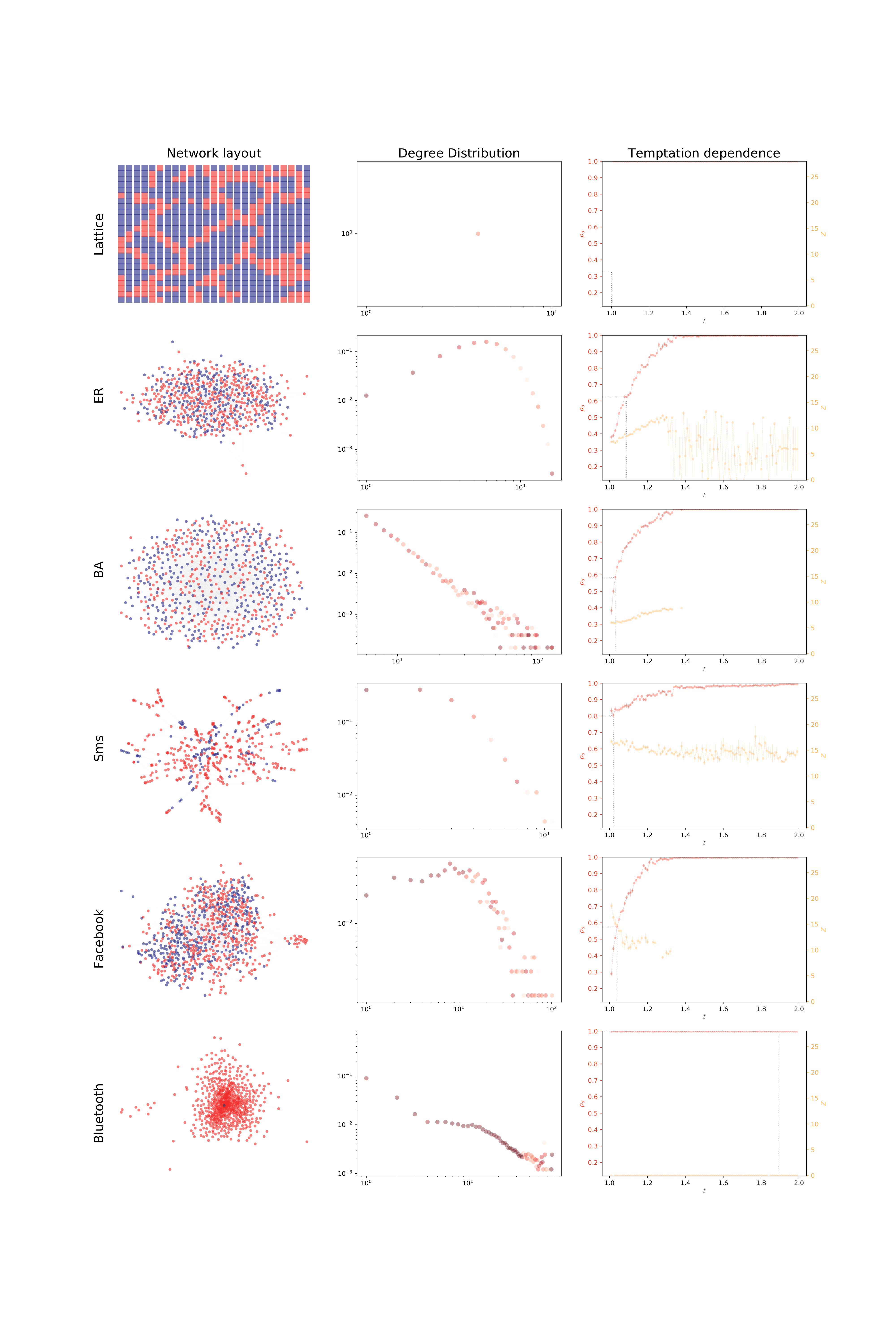}
	\caption{summary of the individual maximum heuristic (\ref{it:heuristic_indmax}).}
	\label{fig:nc_indmax}
\end{figure}

\begin{figure}
	\centering
	\includegraphics[width=\figwidth]{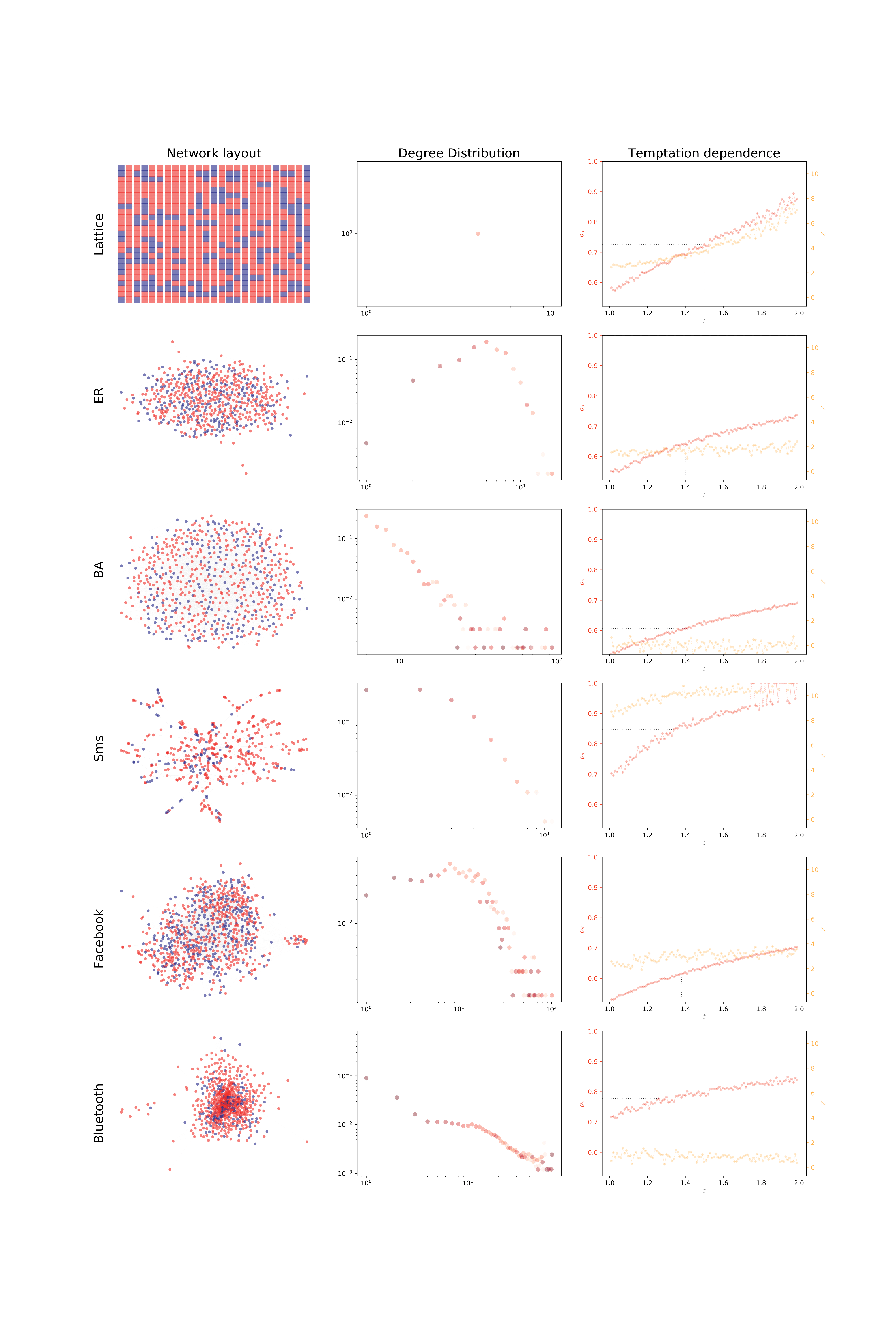}
	\caption{summary of the local softmax heuristic (\ref{it:heuristic_localsoftmax}).}
	\label{fig:nc_localsoftmax}
\end{figure}

\begin{figure}
	\centering
	\includegraphics[width=\figwidth]{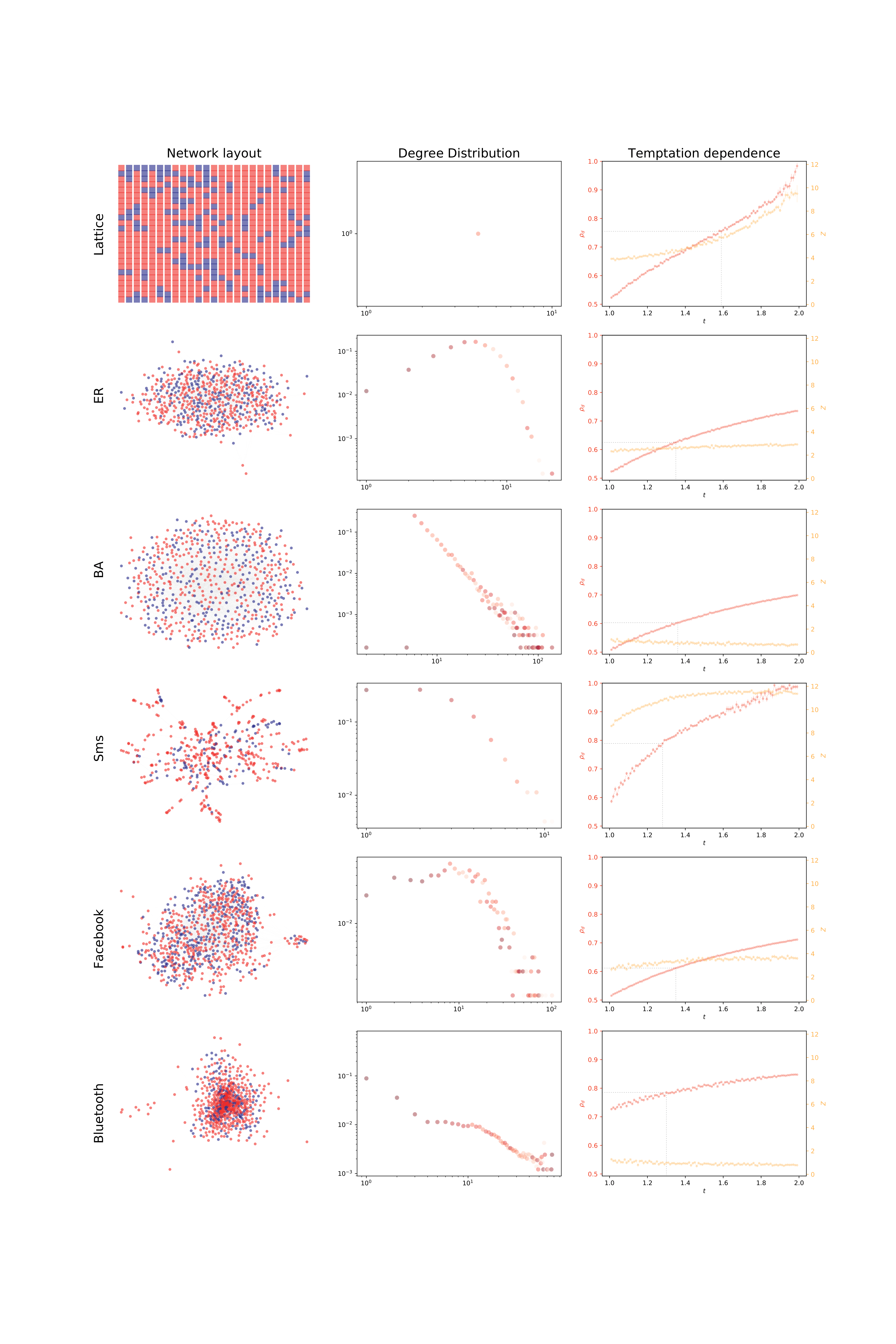}
	\caption{summary of the individual softmax heuristic (\ref{it:heuristic_indsoftmax}).}
	\label{fig:nc_indsoftmax}
\end{figure}

\begin{figure}
	\centering
	\includegraphics[width=\figwidth]{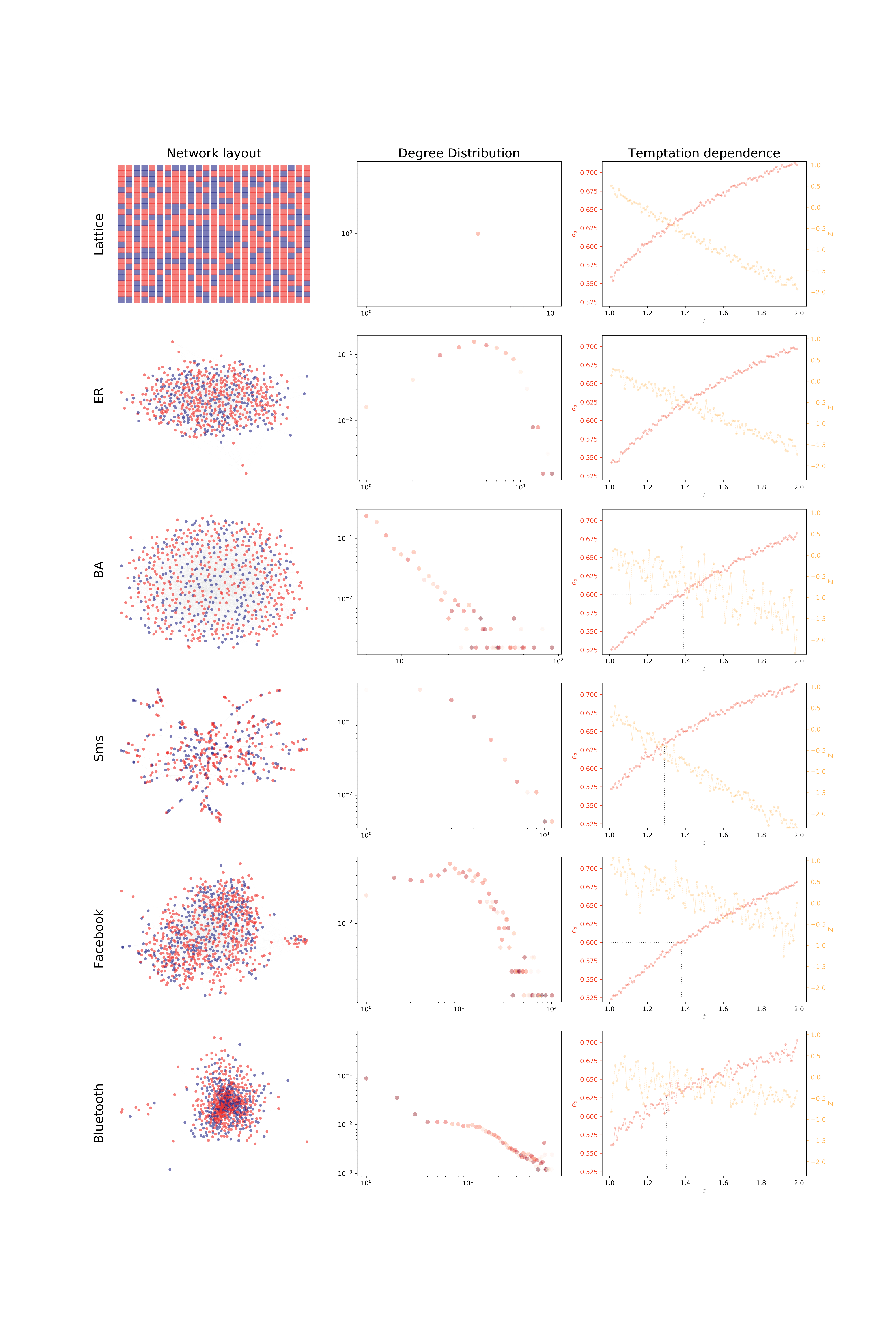}
	\caption{summary of the local stochastic softmax (\ref{it:heuristic_localstocsoftmax}).}
	\label{fig:nc_localstocsoftmax}
\end{figure}

\begin{figure}
	\centering
	\includegraphics[width=\figwidth]{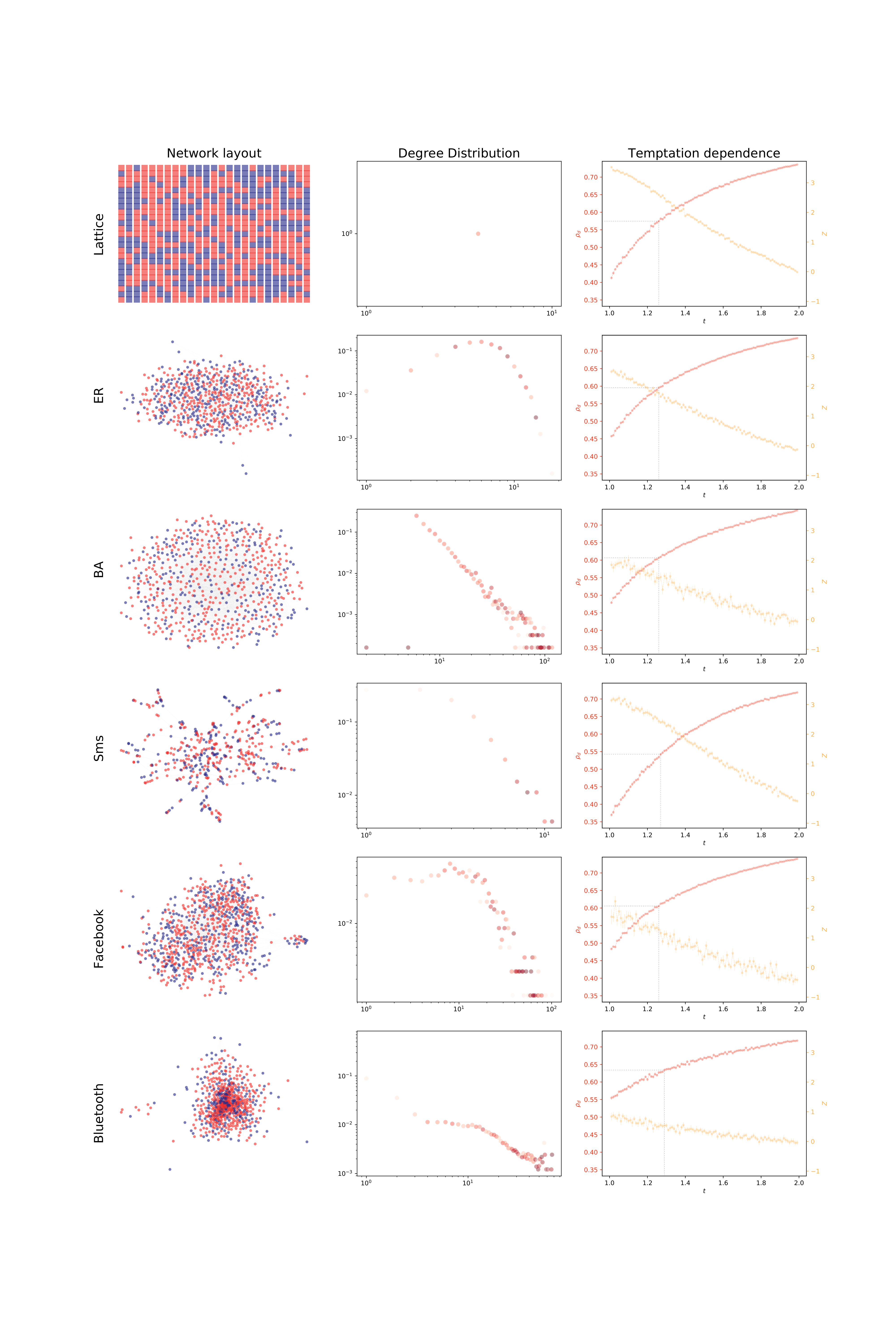}
	\caption{summary of the individual maximum heuristic (\ref{it:heuristic_indstocsoftmax}).}
	\label{fig:nc_indstocsoftmax}
\end{figure}

In addition to this, a range of networks constructed from real-world data were also employed. The data used to construct all such networks comes from the Sensible DTU experiment\cite{Stopczynski2014} at the Technical University of Denmark. The experiment consisted of a large number ($>700$) of Danish university students, who received smartphones which, with their consent, registered information regarding contact patterns, sensor information, etc. The data for a one-month observation period of this study are made publicly available\cite{Sapiezynski2019}. From this, the following networks were constructed.
\begin{itemize}
	\item A text message (SMS) network with any two nodes $(u, v)$ connected if either had texted the other during a one-month observation period, containing a total of $n=457$ nodes.
	\item A Facebook (FB) network  consisting of $n=800$, with users linked if they were friends on Facebook.
	\item A Bluetooth (BT) network with $n=542$ nodes. This network consists of temporal 'slices' of periods of one hour. During each such time slice, an edge $(u, v)$ is present if the corresponding users were in physical proximity of one another during the period, as identified by the Bluetooth sensors in their phones. Proximity was then detected by thresholding signal strengths to RSSI values above $-90dB$, corresponding to distances of a few metres\cite{Sekara2014}.
\end{itemize}
In addition, we considered a range of different update heuristics for agents engaged in repeated games on various graphs. A brief overview of the heuristics considered, as well as some descriptions of their qualitative differences, is provided below.
\begin{enumerate}
	\item \label{it:heuristic_localmax} A 'local maximum' heuristic in which agents consider themselves and their neighbourhood, and copy the strategy of whichever node received the greatest mean payout in the previous round.
	
	\item \label{it:heuristic_indmax} An 'individual max' heuristic, in which agents follow the same procedure as above, but only compare themselves with a single node from their neighbourhood, which they choose uniformly at random.
	
	\item \label{it:heuristic_localsoftmax} An 'local softmax' heuristic, in which nodes consider the payouts earned by themselves in the previous round, and the average payout of neighbors using the opposite strategy. If all neighbors used the same strategy as the nodes, its strategy will not update. Otherwise, it will use the two payouts as inputs to a softmax function which determines the probabilities of the strategies.
	
	\item \label{it:heuristic_indsoftmax} An 'individual softmax' heuristic similar to \ref{it:heuristic_localsoftmax} but comparing the node to a randomly selected neighbor.
	
	\item \label{it:heuristic_localstocsoftmax} A 'local stochastic softmax' heuristic, similar to \ref{it:heuristic_localsoftmax}, but without the constraint that nodes must deterministically reuse their previous strategy if nobody in their neighbourhood played the opposite strategy in the previous round.
	
	\item \label{it:heuristic_indstocsoftmax} A 'individual stochastic softmax' heuristic, similar to \ref{it:heuristic_localstocsoftmax}, but considering only a randomly selected neighbor.
\end{enumerate}
The heuristics outlined in~\cref{it:heuristic_localmax,it:heuristic_indmax,it:heuristic_localsoftmax,it:heuristic_indsoftmax,it:heuristic_localstocsoftmax,it:heuristic_indstocsoftmax} lead to different global dynamics on different networks. In \cref{fig:nc_localmax,fig:nc_indmax,fig:nc_localsoftmax,fig:nc_indsoftmax,fig:nc_localstocsoftmax,fig:nc_indstocsoftmax} these dynamics are shown for a range of network structures. In each figure, each row of subfigures corresponds to a network structure. The right column of subfigures summarise the influence of the 'defection temptation' parameter $t$. The subfigures show the fraction of nodes defecting after $10^4$ simulation steps, averaged over an additional $10^3$ steps (red) as well as the pairing measure (gray) obtained in a similar way. The dashed lines show the value for $t$ resulting in the value of $\rho_I$ that is as close as possible to the midpoint between the maximum and minimum values for $\rho_I$.

For this value $t^*$, an additional simulation was run for each network. The end states after $10^4$ time steps are illustrated in the left columns of subfigures, with red and blue denoting defectors and cooperators, respectively. The central columns show the degree distributions of the networks, coloured based on defection rates - for each degree, a score was computed by taking the average number of times nodes of each degree defected in the last $10^3$ iterations of the simulation. The nodes were then coloured based on the ranks of those scores, with more defection/cooperation corresponding to more blue/red colours.

The above preliminary investigations reproduce several findings from the literature. One example is the sharp phase transition on the square lattice in figure \ref{fig:nc_localmax}, first observed by Nowak and May. Note that the transition occurs at slightly lower values of $t$, because we opted for an asynchronous update scheme in which $10\%$ of nodes, rather than all of them, changed their strategy in each round. Other such findings include under stochastic heuristics, network structure has little to no effect\cite{Gracia-Lazaro2012}, as shown in e.g. \cref{fig:nc_localstocsoftmax}. Finally, we reproduce the finding that, when using a non-deterministic update heuristic, BA and ER networks facilitate similar levels of cooperation\cite{Wu2007}.

In the first experiment - from which the data we intended to fit these heuristics to originated - participants were presented with information about a single node in their vicinity, leading us to limit ourselves to the 'individual' heuristics presented above. The deterministic examples of such heuristics, such as that described in \ref{it:heuristic_indmax} are ill-suited to fit to data, a single data point can have a likelihood of zero. At the same time, we wanted a model for individual behaviour which, like those encountered in the literature, can accommodate relatively stable regions of cooperators and defectors. Hence, we ended up fitting a stochastic individual softmax model like that described in \cref{it:heuristic_indstocsoftmax}, which we then adjusted to have such properties. The methods section of the main paper describes these adjustments as well as the fitting procedure.

\section*{Effects of clustering}
\label{sec:clustering}
In order to investigate the effects of distributing different update heuristics across a network in a non-uniform fashion, we devised a method of sampling from an ensemble of networks with a continuously varying degree of clustering with regards to terminology category, i.e. varying the tendency for nodes to be disproportionally connected to nodes that act according to the same model (meaning model trained on participants exposed to the collectivist, neutral, or individualist terminology).
\subsection*{The clustering parameter \texorpdfstring{$\alpha$}{alpha}}
\begin{figure}
	\centering
	\includegraphics[width=0.9\linewidth]{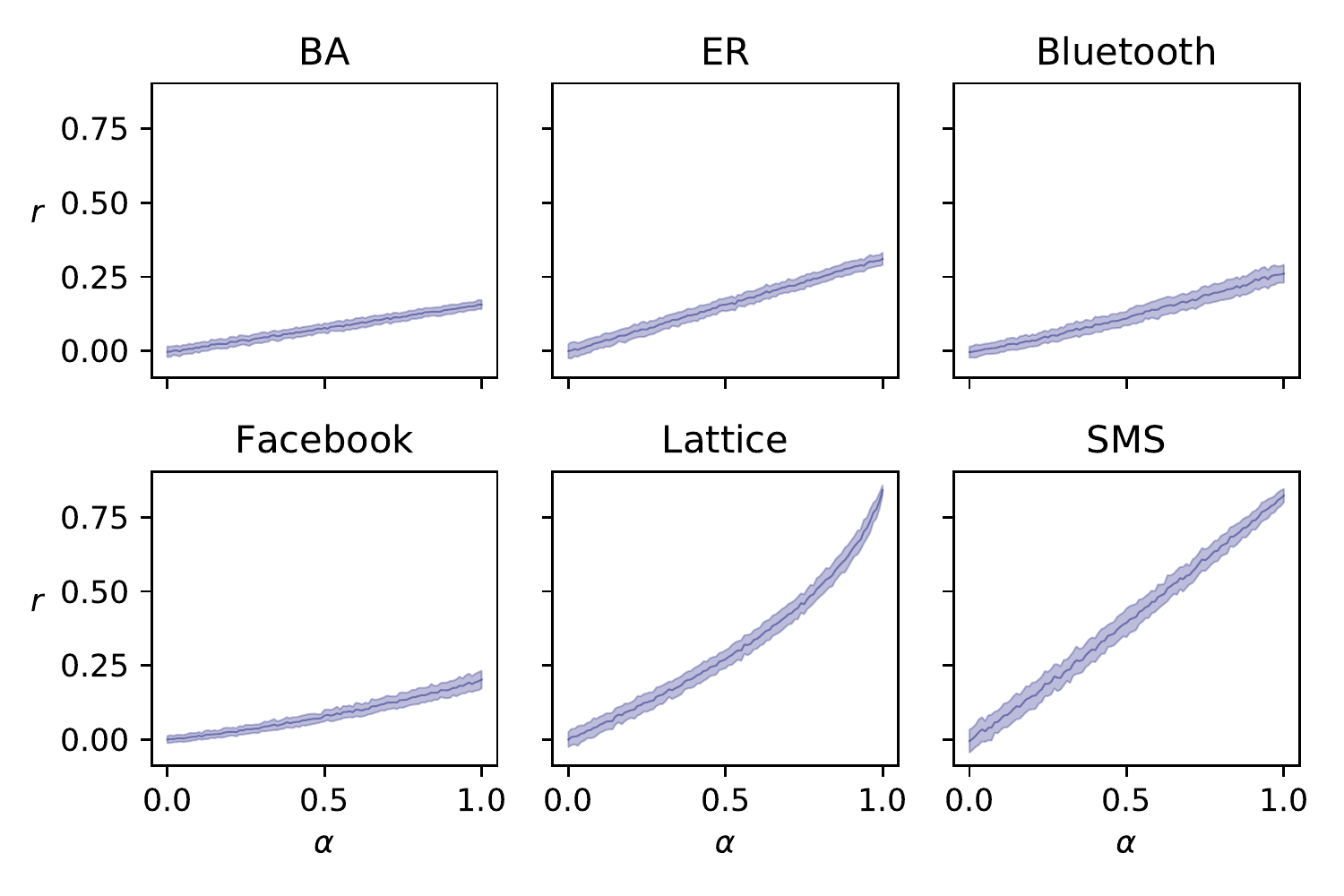}
	\caption{As the clustering hyperparameter $\alpha$ increases, the assortativity coefficient $r$ of the categories grows accordingly. $r$ appears to be growing the most in sparser networks, such as the artificial lattice, or the real text message network.}
	\label{fig:doesalphawork}
\end{figure}

We did this by defining a hyperparameter $\alpha$, signifying the degree of clustering. We then start with a network in which there is no category assigned to the nodes, and perform the assignment in the following fashion. A category is selected based on a predefined parameter $\rho_I$, so categories are chosen with $p_d = \rho_I$ and $p_c = 1 - \rho_I$. Then, with probability $\alpha$, the chosen category $n$ is assigned to a node selected using a Barabási-Albert style preferential attachment mechanism\cite{Barabasi1999}, in which a node $u$ is selected with a probability proportional to the number of its neighbours having been assigned category $n$. With probability $1 - \alpha$, the node is selected uniformly at random. This procedure is repeated until each node has been assigned a category. Figure \ref{fig:doesalphawork} shows the dependence of the category assortativity coefficient $r$ on $\alpha$ for a range of network structures.

\subsection*{Simulation results}
\label{sec:norm_matrices}
In the following, we present the results of a series of simulations and show the resulting metrics discussed in \cref{sec:clustering}. For each metric, we run a series of simulations, for a range of values of $\rho_I$ and $t$, and for three values of the clustering parameter $\alpha$, repeated for the SMS, FB, BA, and SL networks discussed in \cref{sec:heuristics_and_networks}. For each combination of network and $\alpha$ value, we present a 2D heatmap showing how the quantity in question changes with $\rho_I$ and $t$. Each cell in these matrices is computed by running $10$ simulations for $10^4$ iterations, and averaging the values taken by the quantity over the last 5000 iterations. The cell is left white in cases where the measure is ill-defined - for example, the pairing measure is not defined when all nodes either cooperate or defect, as the denominator becomes zero in those cases.

\begin{figure}
	\centering
	\includegraphics[width=\textwidth]{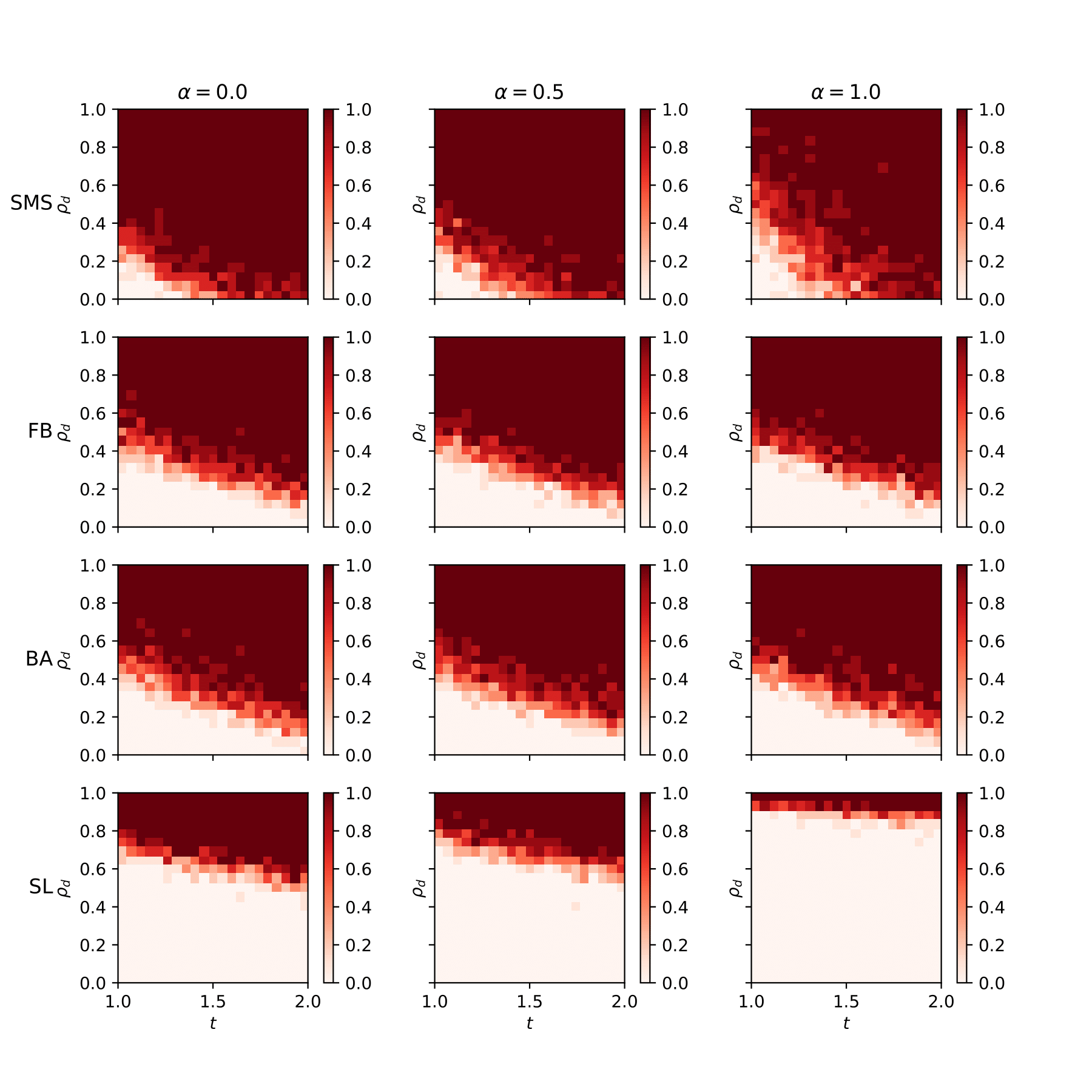}
	\caption{The fraction of defectors in the network after running the simulations.}
	\label{fig:fraction_defecting}
\end{figure}

\begin{figure}
	\centering
	\includegraphics[width=\textwidth]{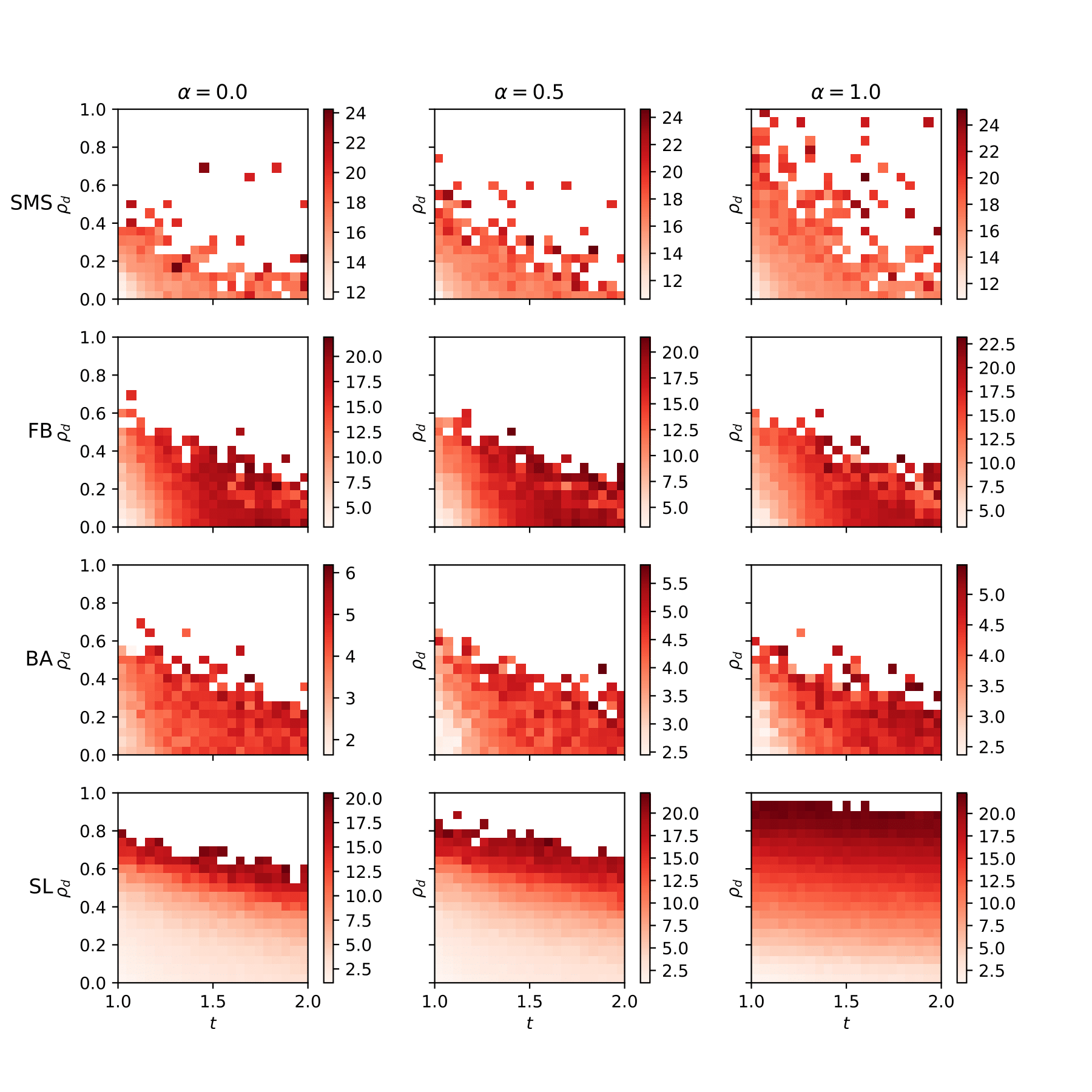}
	\caption{The prevalence measure $Z_c$.}
	\label{fig:pairing_measure}
\end{figure}

\end{document}